\documentclass[pra,twocolumn,superscriptaddress,floatfix,showpacs,aps,letterpaper,nobalancelastpage]{revtex4-2}

\usepackage[braket, qm]{qcircuit}

\usepackage{graphicx} 
\usepackage{color} 
\usepackage{transparent}
\usepackage{amsmath}
\usepackage{hyperref} 
\usepackage{amssymb}
\usepackage{amsthm}
\usepackage{natbib}
\usepackage{qcircuit}
\usepackage{tabulary}
\usepackage[caption=false]{subfig}

\usepackage{pgfplots}
\usepackage{siunitx}
\usepackage{comment}
\usepackage{tikz}
\usepackage[compat=newest]{yquant}

\newcommand{\proj}[1]{|#1\rangle\langle#1|}
\newcommand{\dyad}[2]{|#1\rangle\langle#2|}
\newcommand{\Tr}{{\rm Tr}}
\newcommand{\mte}[2]{\langle#1|#2|#1\rangle }
\newcommand{\ot}{\otimes }
\newcommand{\expc}[2]{\langle#1\rangle_{#2} }

\newcommand{\Ibb}{\mathbb{I}}
\newcommand{\rB}{\textbf{r}}
\newcommand{\pB}{\textbf{p}}
\newcommand{\qB}{\textbf{q}}

\newcommand{\DC}{{\mathcal D}}
\newcommand{\NC}{{\mathcal N}}
\newcommand{\MC}{{\mathcal M}}
\newcommand{\IC}{{\mathcal I}}
\newcommand{\HC}{{\mathcal H}}
\newcommand{\PC}{{\mathcal P}}

\newcommand{\ep}{\epsilon}
\newcommand{\lm}{\lambda }
\newcommand{\Lm}{\Lambda }

\renewcommand{\th}{\theta } 

\newcommand{\new}[1]{\textcolor{black}{#1}}
\usepackage[normalem]{ulem}

\begin{document}

\title{Basic \new{entanglement} distillation with realistic noise}

\author{Vikesh Siddhu}
\author{Erick Winston}
\author{David C. McKay}
\email{dcmckay@us.ibm.com}
\author{Ali Javadi-Abhari}
\email{ali.javadi@ibm.com}
\affiliation{IBM Quantum, IBM T.J. Watson Research Center, Yorktown Heights, NY 10598, USA}

\begin{abstract}
    Entanglement distillation is a key component of modular quantum computing
    and long-range quantum communications. However, this powerful tool to
    reduce noise in entangled states is difficult to realize in practice for
    two main reasons.  First, operations used to carry out distillation inject
    noise they seek to remove.  Second, the extent to which distillation can
    work under realistic device noise is less well-studied.  In this work, we
    both simulate distillation using a variety of device noise models and
    perform distillation experiments on fixed-frequency IBM devices.  We find
    reasonable agreement between experimental data and simulation done using
    Pauli and non-Pauli noise models.  In our data we find broad improvement
    when the metric of success for distillation is to improve average Bell
    fidelity under effective global depolarizing noise, or remove coherent
    errors, or improve the Bell fidelity of mildly degraded Bell pairs.  We
    pave the way to obtain broad improvement from distillation under a
    stricter, but practically relevant, metric: distill \new{(physically)}~non-local Bell pairs
    with higher fidelity than possible to obtain with other available methods.
    Our results also help understand metrics and requirements for quantum
    devices to use entanglement distillation as a primitive for modular
    computing. 
\end{abstract}

\maketitle

\section{Introduction}

Entanglement is an important non-classical resource used in quantum computation
and communication. However, this resource depletes due to noise that inevitably
affects physical systems. This noise may be removed using entanglement
distillation~(also called
purification)~\cite{BennettBrassardEA96,BennettDiVincenzoEA96}. At a high
level, entanglement distillation entangles many copies of lower fidelity Bell
states such that we can post-select on a subspace to obtain higher fidelity
Bell pairs. Therefore, this basic technique generally requires a high
throughput of Bell pairs and its simplest variants are non-deterministic.
However, via enough rounds of distillation and consumption of Bell pairs, one
can hope to achieve high fidelities~\cite{DurBriegelEA99}~(see also recent
work~\cite{Gidney2023}). In the context of quantum error
correction~(QEC)~\cite{Shor95, Knill05} and quantum
compilation~\cite{BravyiEA16}, high fidelity Bell pairs can enable measurement
and (gate)~teleportation.  When passing states through certain channels,
distillation can even outperform QEC for
communication~\cite{BennettDiVincenzoEA96}.

There is a vast body of work dedicated to entanglement
distillation~\cite{KimYunEA24, ShiLiuEA24, YanZhouEA23, KangGuhaEA23,
MiguelRamiroRieraSabatEA23, VandreGuhne23, DevulapalliSchouteEA22,
JansenGoodenoughEA22, MiguelRamiroDur18, FujiiYamamoto09, GlancyKnillEA06,
AschauerBriegel02, PanSimonEA01, PattisonBaranesEA24, YuPatilGuha25,
YuPatilGuha25A}.  From a theoretical perspective, understanding the best rates
for distilling entanglement and methods for achieving these rates in practice
is an important area of study. In practice, entanglement distillation plays a
central role in both long-distance quantum communication~\cite{DurBriegelEA99,
BriegelDurEA98a, HuHuangEA21} and scaling quantum computation, for instance
using modular architectures~\cite{SteinSussmanEA23, AngCariniEA22}. A variety
of studies aim at optimizing protocols in these
settings~\cite{RozpedekSchietEA18, KrastanovAlbertEA19, VictoraTserkisEA23,
ZhaoZhaoEA21,ZangEA25}.

Due to its importance, a variety of prior experiments on various physical
platforms including optical~\cite{Kwiat2001, PanGasparoniEA03, Pan2001,
Pan2003, Yamamoto2003, Dong2008, Chen2017, HuEA21}, trapped
ion~\cite{ReichleLeibfriedEA06}, solid state~\cite{KalbReisererEA17}    , and
superconducting ~\cite{YanZhongEA22} qubits have reported entanglement
distillation.  Such proof-of-concept studies have opened the way for using
entanglement distillation not only in quantum networks but also modular
architectures for quantum computing. However, a variety of challenges exist in
understanding the role distillation can play in practice, both in-terms of
finding appropriate distillation protocols that can be implemented, and the
efficacy of using distillation protocols over other strategies for distributing
Bell pairs given the noise in the hardware components implementing these
protocols.

Carrying out distillation on devices can be non-trivial and analyzing it
requires care for noise sources different from those considered in a large part
of entanglement distillation literature. For instance, a number of theoretical
studies assume that the entangled states experience independent identically
distributed~(iid) noise~(typically Pauli noise). Furthermore, many assume
noiseless gates, measurements, and ancilla. In practice, entangled states
experience non-iid noise. For instance, the natural way to create two
\new{physically} non-local Bell pairs~\new{(a physically non-local Bell pair
refers to a Bell state on physically unconnected qubits.)} may add noise
differently to each pair~\cite{QinDuEA23}. This noise need not be Pauli
noise~\cite{AliferisBritoEA09}, for instance most qubits have a non-negligible
$T_1$ time, which is a non-Pauli error. The gates used to carry out
entanglement distillation also add noise. Measurements done during or after
distillation are also noisy and, typically, of long duration. These measurement
errors not only affect the distillation protocol, but also the protocols used
to certify how well distillation works.  Finally, error-free ancillas, used in
certain nested schemes for distillation, need not be available either.

In this work, we numerically and experimentally explore entanglement
distillation on IBM's superconducting qubits. In our numerical exploration we
include noise on both the Bell pairs being distilled and the
components~(two-qubit gates and measurements) carrying out this distillation.
For local Pauli noise we report gate and measurement noise parameters for three
different distillation protocols to improve the Bell fidelity of noisy Bell
pairs~(with possibly unequal initial Bell fidelity). For global depolarizing
noise on the Bell pairs we also report experimental results that we fit
numerically. In further experimental exploration we report the performance of
three different distillation protocols as a function of the input Bell fidelity
which we degrade by idling the Bell states before distillation. We model the
waiting time error using a non-Pauli noise model~(using the damping-dephasing
channel~\cite{AliferisBritoEA09, SiddhuAbdelhadiEA24}) and include $ZZ$
crosstalk to obtain reasonable agreement with experimental data both when
distillation provides an improvement in Bell fidelity and when it does not. 

Our results show that (1) simple noise parameters of a device can be turned
into a non-Pauli noise model to capture essential features of our entanglement
distillation protocols; (2) using the noise model it is possible to pick one
distillation protocol in favor of another; and (3) depending on the metric of
success for distillation, one may or may not find broad improvement from the
simplest distillation protocols. 

The rest of this paper is organized as follows. Using standard notation~(see
App.~\ref{sec:notation}), we first summarize well-known ideas of entanglement
distillation in Sec.~\ref{sec:distill}. These ideas include the recurrence
protocol, what we call the $\{ZX_{3B}\}$ distillation protocol~(also called
double selection in~\cite{FujiiYamamoto09}), impact of distillation on Bell
pairs of unequal Bell fidelity, and distillation under global depolarizing
noise. The next section, Sec.~\ref{sec:localDepol}, is devoted to studying the
effects of gate noise, measurement noise, and local depolarizing noise on Bell
state preparation and entanglement distillation. In Sec.~\ref{sec:globalDepol}
we augment the noise model of the previous section by adding global
depolarizing noise to qubits as they wait prior to being distilled. We generate
experimental data on superconducting qubits to mimic depolarizing noise using
twirling.  Finally, in Sec.~\ref{sec:deviceNoise} we report results on the
experimental study of three distillation protocols, two recurrence-type
protocols and the $ZX_{3B}$ protocol, as a function of idling noise on
superconducting qubits; we match this data numerically using a
damping-dephasing~(non-Pauli) noise model also described in that section.

\section{Entanglement Distillation}
\label{sec:distill}

In this section we summarize the concept of distillation based on and extending
the recurrence protocol~\cite{BennettBrassardEA96,BennettDiVincenzoEA96} and
show some simple calculations based on depolarizing models focusing on how the
protocols suffer if there is an imbalance in the fidelities of the input
states. \new{Here fidelity refers to what we call the Bell fidelity, $F :=
\mte{\phi}{\rho}$, between a two-qubit density operator $\rho$ and a fixed
maximally entangled state $\ket{\phi} = (\ket{00} + \ket{11})/\sqrt{2}$.}

\subsection{Recurrence ($Z_{2B}$, $X_{2B}$)}
\label{sec:rec}

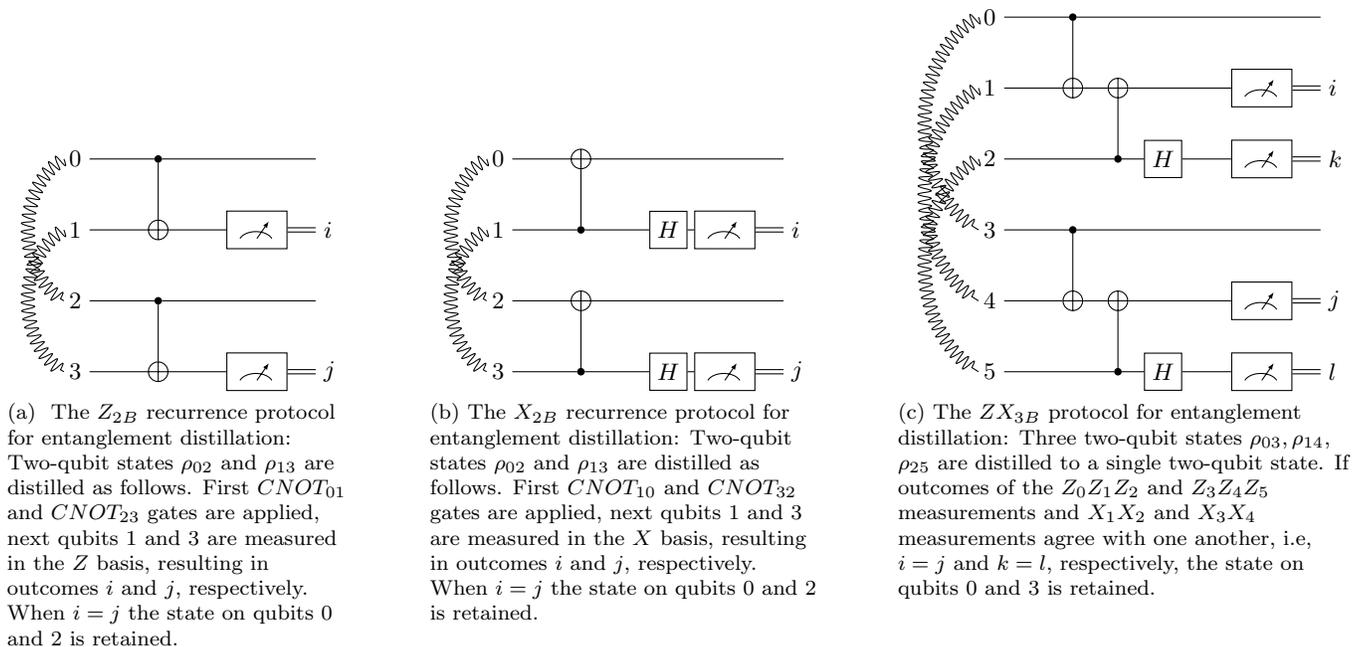
\begin{figure*}[htp]
     \subfloat[ \raggedright The $Z_{2B}$ recurrence protocol for entanglement
     distillation: Two-qubit states $\rho_{02}$ and $\rho_{13}$ are distilled
     as follows. 
    First $CNOT_{01}$ and $CNOT_{23}$ gates are applied, next qubits $1$
    and $3$ are measured in the $Z$ basis, resulting in outcomes $i$ and $j$,
    respectively.
    When $i=j$ the state on qubits $0$ and $2$ is retained.
    ]
     {
    \begin{tikzpicture}
    \begin{yquant}[register/minimum height=12pt, register/minimum depth=12pt]
    [name=bell11] qubit  {$0$} b11;
    [name=bell21] qubit  {$1$} b21; 
    [name=bell12] qubit  {$2$} b12; 
    [name=bell22] qubit  {$3$} b22; 
    \draw[decorate,decoration={coil,aspect=0,segment length=3pt}] (bell11.west) to [bend right=60]  (bell12.west) ;
    \draw[decorate,decoration={coil,aspect=0,segment length=3pt}] (bell21.west) to [bend right=60]  (bell22.west) ;
    hspace {16pt} -;
    cnot b21 | b11;
    cnot b22 | b12;
    hspace {16pt} -;
    measure b21;
    measure b22;
    hspace {8pt} -;
    output {$i$} b21;
    output {$j$} b22;
    \end{yquant}
    \end{tikzpicture}
    \label{fig:distillBasicZ2B}
    }
    \hfill
     \subfloat[\raggedright The $X_{2B}$ recurrence protocol for entanglement distillation: Two-qubit
    states $\rho_{02}$ and $\rho_{13}$ are distilled as follows. 
    First $CNOT_{10}$ and $CNOT_{32}$ gates are applied, next qubits $1$
    and $3$ are measured in the $X$ basis, resulting in outcomes $i$ and $j$,
    respectively.
    When $i=j$ the state on qubits $0$ and $2$ is retained.
     ]{
    \begin{tikzpicture}
    \begin{yquant}[register/minimum height=12pt, register/minimum depth=12pt]
    [name=bell11] qubit  {$0$} b11;
    [name=bell21] qubit  {$1$} b21; 
    [name=bell12] qubit  {$2$} b12; 
    [name=bell22] qubit  {$3$} b22; 
    \draw[decorate,decoration={coil,aspect=0,segment length=3pt}] (bell11.west) to [bend right=60]  (bell12.west) ;
    \draw[decorate,decoration={coil,aspect=0,segment length=3pt}] (bell21.west) to [bend right=60]  (bell22.west) ;
    hspace {16pt} -;
    cnot b11 | b21;
    cnot b12 | b22;
    hspace {16pt} -;
    h b21;
    measure b21;
    h b22;
    measure b22;
    hspace {8pt} -;
    output {$i$} b21;
    output {$j$} b22;
    \end{yquant}
    \end{tikzpicture}
    \label{fig:distillBasicX2B}
}
    \hfill
    \subfloat[\raggedright The $ZX_{3B}$ protocol for entanglement distillation: Three
    two-qubit states $\rho_{03}, \rho_{14}$, $\rho_{25}$ are distilled to a
    single two-qubit state. If outcomes of the $Z_0Z_1Z_2$ and $Z_3Z_4Z_5$
    measurements and $X_1X_2$  and $X_3X_4$ measurements agree with one
    another, i.e, $i=j$ and $k=l$, respectively, the state on qubits $0$ and
    $3$ is retained.]{
    \begin{tikzpicture}
    \begin{yquant}[register/minimum height=12pt, register/minimum depth=12pt]
    [name=bell11] qubit  {$0$} b11;
    [name=bell21] qubit  {$1$} b21; 
    [name=bell31] qubit  {$2$} b31;
    [name=bell12] qubit  {$3$} b12; 
    [name=bell22] qubit  {$4$} b22; 
    [name=bell32] qubit  {$5$} b32; 

    \draw[decorate,decoration={coil,aspect=0,segment length=3pt}] (bell11.west) to [bend right=60]  (bell12.west) ;
    \draw[decorate,decoration={coil,aspect=0,segment length=3pt}] (bell21.west) to [bend right=60]  (bell22.west) ;
    \draw[decorate,decoration={coil,aspect=0,segment length=3pt}] (bell31.west) to [bend right=60]  (bell32.west) ;
    hspace {16pt} -;
    cnot b21 | b11;
    cnot b22 | b12;
    cnot b21 | b31;
    cnot b22 | b32;
    h  b31;
    h  b32;
    hspace {16pt} -;
    measure b21;
    measure b22;
    measure b31;
    measure b32;
    hspace {8pt} -;
    output {$i$} b21;
    output {$j$} b22;
    output {$k$} b31;
    output {$l$} b32;
    \end{yquant}
    \end{tikzpicture}
    \label{fig:distill3To1}
     }
     \caption{Basic circuits for entanglement distillation}
     \label{fig:distillBasic}
\end{figure*}

The minimal recurrence protocols, using two Bell states (four qubits, $0,1,2,$
and $3$), are shown in Fig.~\ref{fig:distillBasicZ2B} and
Fig.~\ref{fig:distillBasicX2B}. In Fig.~\ref{fig:distillBasicZ2B} we show
$Z_{2B}$ recurrence where the circuit measures $Z_0 Z_1$ and $Z_2 Z_3$, and
post-selects for the case where these two measurements, $i$ and $j$,
respectively, equal each other.  Replacing these $ZZ$ measurements with $XX$
measurements results in what we call $X_{2B}$ recurrence, shown in
Fig.~\ref{fig:distillBasicX2B}.

Due to the nature of the measurement, $Z_{2B}$ can detect single $X$ or $Y$
errors and $X_{2B}$ can detect single $Z$ or $Y$ errors. In essence, these
protocols measure a stabilizer of the Bell state (either $ZZ$ or $XX$) and
discard upon encountering an error. The only difference between Bell state
distillation compared to other forms of error detection in Clifford circuits is
that the stabilizer measurement happens in a distributed fashion, i.e. instead
of computing the parity of the Bell state into one qubit, we compute it into an
entangled resource. This allows the parity checks to happen locally on each
half of the Bell pair without further (quantum) communication across different
halves of a Bell pair.

We can gain some intuition for the performance of the protocol by applying
$Z_{2B}$ to Bell states with a stochastic bit flip, 
\begin{equation}
    \rho_{02} = \IC \ot \DC_{p}(\phi_{02}) \quad \text{and}
    \quad \rho_{13} = \IC \ot \DC_q(\phi_{13}),
    \label{eq:noiseChannel}
\end{equation}
where $\phi$ is the maximally entangled state, the identity channel $\IC$ acts
on qubits $0$ and $1$ and the bit flip channels $\DC_p$ and $\DC_q$~(see
App.~\ref{sec:notation}) act on qubits 2 and 3, respectively, with bit flip
probabilities $p$ and $q$, \new{respectively}. The measurements $i$ and $j$ in
Fig.~\ref{fig:distillBasicX2B} equal each other with probability $p_s =
(1-p)(1-q) + pq$~(sometimes called the acceptance probability \new{or
acceptance fraction}) and result in
a distilled state $\rho_{02}' = \IC \ot \DC_{r}(\phi)$ where $r = pq/p_s$. The
maximum Bell fidelity before distillation, and Bell fidelity after distillation
take values,
\begin{equation}
    F_b = \max(1 - p,  1 - q),  \quad \text{and} \quad
    F_a = \frac{(1-p)(1-q)}{p_s},
    \label{eq:FidDistill}
\end{equation}
respectively. Interchanging $p$ and $q$ does not change $F_a, F_b$, and $p_s$,
thus we let $p \leq q$, and write
\begin{equation}
    F_a - F_b = \frac{1}{p_s}p(1-p)(1-2q) \geq 0,
    \label{eq:incFid}
\end{equation}
where the inequality is strict whenever $0 < p \leq q < 1/2$. The strict
inequality implies distillation improves Bell fidelity for all non-trivial $p$
and $q$ (under the assumption of a perfect implementation of the distillation
protocol). 
This example demonstrates (1) recurrence in Fig.~\ref{fig:distillBasicZ2B}
reduces the $X$ error rate from linear~($O(p)$ or $O(q)$) to quadratic,
$O(pq)$, with probability $p_s$; (2) the reduction occurs by post-selecting
away those errors on qubits $2$ and $3$ which anti-commute with the $Z_2Z_3$
measurement.  Similar conclusions hold for the $X_{2B}$ recurrence protocol
when analyzed for phase-flip noise, i.e., $Z$ errors.

\subsection{$ZX_{3B}$ Distillation}
\label{sec:ZZZIXX}

The limitation in the previous recurrence protocols was that they are not
simultaneously sensitive to phase flip and bit flip error. If we increase the
number of input Bell states to three (six qubits), then we can perform a
distillation protocol with additional checks~\cite{FujiiYamamoto09}. The
circuit for the protocol is given in Fig.~\ref{fig:distill3To1}. One way to
think about the protocol is to replace the $ZZ$ measurements in recurrence with
two sets of measurements, one set measuring $ZZZ$ and another set measuring
$IXX$~(notice $ZZZ$ and $IXX$ commute with each other and thus can be measured
simultaneously). Like recurrence, one accepts the final state only when
measurement outcomes agree. A more general treatment of distillation protocols
based on simultaneous measurement of commuting observable is available in
App.~\ref{sec:glnDist}

\subsection{Distilling depolarized qubits}
\label{sec:dist_depol_1}

\begin{figure*}[htp]
    \subfloat[\raggedright Increase in Bell fidelity, see eq.~\eqref{eq:incFid},
    using $Z_{2B}$ distillation on Bell states undergoing bit flip
    errors~\eqref{eq:noiseChannel}.]
    {
        \includegraphics[scale=.4]{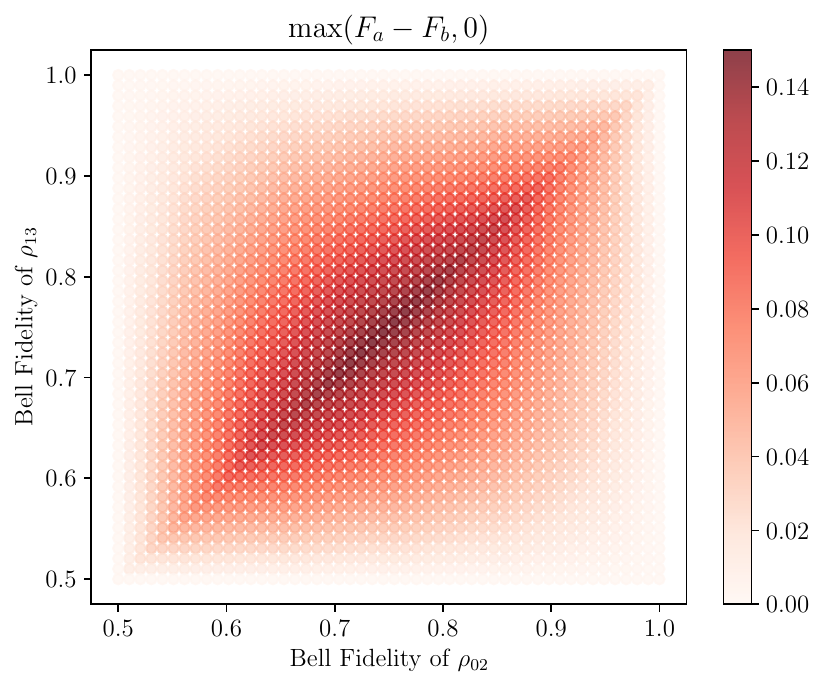}
        \label{fig:xDephDiff}
    }
    \hfill
     \subfloat[\raggedright Increase in Bell fidelity, see
     eq.~\eqref{eq:2To1FidDistill}, using $Z_{2B}$ distillation on Bell states
     undergoing depolarizing noise~\eqref{eq:2To1noiseChannel}.]
     {
         \includegraphics[scale=.4]{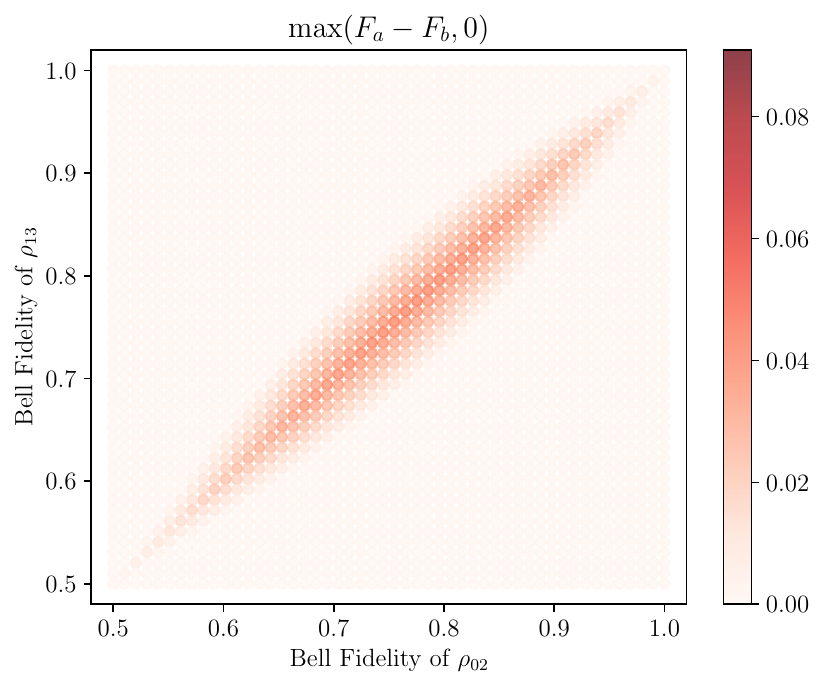}
        \label{fig:depolUneq}
    }
    \hfill
    \subfloat[][\raggedright Increase in Bell fidelity, see
    eq.~\eqref{eq:3To1FidDistill}, using $ZX_{3B}$ distillation on Bell states
    undergoing depolarizing noise~\eqref{eq:3To1noiseChannel}.]
    {
        \includegraphics[scale=.4]{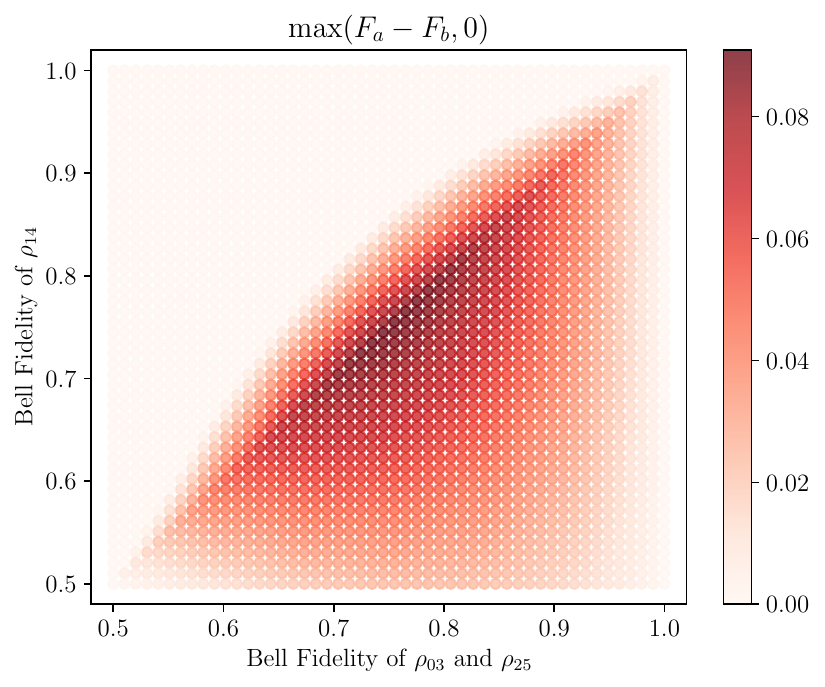} 
        \label{fig:3To1DepolUneq}
    }
    \caption{Algebraically obtained increase in Bell fidelity by different
    distillation  protocols. The $x$ and $y$ axes represent the Bell fidelity
    of the states prior to distillation while the color corresponds to the
    increase~(if any) in fidelity upon distillation.}
\end{figure*}

To better understand the performance of these distillation protocols for
general noise, we look at these protocols when the Bell pairs are depolarized.
In Fig.~\ref{fig:distillBasicZ2B} suppose the Bell pairs are acted on by local
depolarizing channels $\Lambda$~(see App.~\ref{sec:notation} for notation),
\begin{equation}
    \rho_{02} = \IC \ot \Lm_{p}(\rho) \quad \text{and} \quad
    \rho_{13} = \IC \ot \Lm_{q}(\rho)
    \label{eq:2To1noiseChannel}
\end{equation}
with probabilities $p$ and $q$, respectively. Then the before and after
distillation Bell fidelities are 
\begin{equation}
    F_b = \max(1-p, 1-q), \quad  \text{and} \quad
    F_a = \frac{1}{p_s}\big((1-p)(1-q) + \frac{pq}{9} \big),
    \label{eq:2To1FidDistill}
\end{equation}
where 
\begin{equation}
    p_s = (1 - 2p/3)(1-2q/3) + 4pq/9,
    \label{eq:pAcc2To1}
\end{equation}
is the acceptance probability~(see App.~\ref{ap:distCal} for details). In
contrast to~\eqref{eq:incFid} where $F_a$ is generally larger than $F_b$, here
$F_a$ is larger than $F_b$ for a smaller set of initial Bell fidelities; in
Fig.~\ref{fig:depolUneq} we highlight this region where $F_a > F_b$ for roughly
$.19$ fraction of the points sampled.

If we distill three Bell pairs (using the $ZX_{3B}$ protocol) there is a
comparatively larger region where the protocol shows improvements. In
particular, assume the input to the circuit shown in
Fig.~\ref{fig:3To1DepolUneq} are Bell pairs acted on by local depolarizing
channels, 
\begin{equation}
    \rho_{03} = \rho_{25} = \IC \ot \Lm_{p}(\rho) \quad \text{and} \quad
    \rho_{14} = \IC \ot \Lm_{q}(\rho),
    \label{eq:3To1noiseChannel}
\end{equation}
where we assume two of the pairs have equal noise. Before and after
distillation the Bell fidelities are, 
\begin{align}
    \begin{aligned}
        F_b &= \max(1-p, 1-q), \quad \text{and} \\
        F_a &= \frac{1}{p_s} 
        \big( p^2(1-\frac{28}{27}q) + p(\frac{19}{9}q - 2) + 1 - q \big),
    \end{aligned}
    \label{eq:3To1FidDistill}
\end{align}
and the acceptance probability is 
\begin{equation}
    p_s = \frac{p^2}{9}\big(8 - \frac{32}{3}q\big) + \frac{p}{3} \big( \frac{20}{3} q - 5 \big) + 1-q.
    \label{eq:3To1ps}
\end{equation}
The set of initial Bell fidelities where $F_b > F_a$ is given in
Fig.~\ref{fig:3To1DepolUneq}. In comparison, this region~($\simeq .61$
fraction of points) is about three times larger than the one in
Fig.~\ref{fig:depolUneq}.

\subsection{Global depolarizing noise}

Next we consider the effect of the protocols when the Bell pairs are subject to
an $n$-qubit (global)~depolarizing channel,
\begin{equation}
    \NC_{\lm}(\rho) = (1-\lm) \rho + \lm \Tr(\rho) \frac{\Ibb_{2^n}}{2^n}.
\end{equation}
\new{The channel is simple to analyze, represents correlated noise among the
qubits it acts on and can be simulated in experiments~(see
Sec.~\ref{sec:glDplExp}).} We can use the formalism described in
Appendix~\ref{sec:glnDist} to obtain an acceptance probability,
\begin{equation}
    p_G = (1-\lm)p_a + \frac{\lm}{2^{n-1}}
\end{equation}
and Bell fidelity after distillation, 
\begin{equation}
    F_G = \frac{1}{p_G} \big( (1 - \lm) F p_a + \frac{\lm}{2^{n+1}} \big),
\end{equation}
where $p_a$ and $F$ are given in~\eqref{eq:pAccGln} and~\eqref{eq:postSelF}
respectively.  If $\rho_{AB} = \phi^{\ot 2}$ and we use the $Z_{2B}$
protocol~(see Sec.~\ref{sec:rec}) for distillation, then $p_a = 1$
in~\eqref{eq:pAccGln}, $F = 1$ in~\eqref{eq:postSelF}, fidelity of the state
before distillation is $F_b = 1 - 3\lm/4$, the acceptance probability, 
and Bell fidelity after distillation take values,
\begin{equation}
    p_G = 1 - \frac{1}{2}\lm, \quad \text{and} \quad
    F_G = \frac{1}{p_G}(1 - \frac{7}{8}\lm),
    \label{eq:RecurGlobal}
\end{equation}
respectively. Notice $r:= F_G/F_b > 1$ for all $0 < \lm < 1$, i.e.,
distillation always improves the Bell fidelity for global depolarizing noise.
This same conclusion holds when $\rho_{AB} = \phi^{\ot 3}$ and the $ZX_{3B}$
distillation protocol is applied. In this case,
\begin{equation}
    p_G = 1 - \frac{3}{4}\lm, \quad \text{and} \quad
    F_G = \frac{1}{p_G}(1 - \frac{15}{16}\lm).
\end{equation}

\section{Circuit noise}

Unlike the previous section in this section we relax the assumption that the
circuit used for distillation is perfect and include gate noise and measurement
error. The addition of these will greatly affect the ability to perform
successful distillation.  Motivated by the planar connectivity of
superconducting qubits (and connecting to our experiments in the later
sections), we consider circuits where Bell pairs need to be created locally and
swapped so that the distillation circuit can be performed; see
Fig.~\ref{fig:TwoToOne} as an example.

To model noise on measurements we apply a bit flip channel, $\DC_m$, prior to
measurement and noise on two-qubit gates is modelled by adding a two-qubit
(global)~depolarizing channel, $\NC_g$, on the two qubits involved in the gate.
Other sources of noise, such as imperfections in initializing the qubits to
$\ket{0}$ and those in implementing single qubit gates, are ignored since these
sources of noise can be comparatively smaller than measurement and two qubit
gate noise. In the following, we will consider distillation as a function of
these noise parameters versus the fidelity of the input Bell pairs. 

\subsection{Local Depolarizing noise}
\label{sec:localDepol}

In the studies presented in this we section we vary the input Bell pair's
fidelity using local depolarizing channels similar to
Sec.~\ref{sec:dist_depol_1}. In Fig.~\ref{fig:circsDist} we show circuits for
distilling Bell pairs under the noise model just described.  Here, we add a
qubit depolarizing channel in two places.  These two places are before the
first barrier~(dotted vertical line) and between the second and third barriers
in each of the sub-figures of Fig.~\ref{fig:circsDist}.  \new{Between each of
these barriers, depolarizing noise is added to only one half of a Bell pair
since noise on both halves of a Bell pair can be transferred to just one using
the standard transpose trick \footnote{The transposition trick tranfers any
single qubit operator $A$ acting on one half of a Bell pair $\ket{\phi}$ to the
other half using using the identity $\Ibb \ot A \ket{\phi} = A^T \ot \Ibb
\ket{\phi}$ }}. This allows us to independently control the asymmetry among the
Bell pairs and fidelity of the Bell pairs prior to distillation. 

\subsubsection{Recurrence}
\label{sec:Local2To1}

\begin{figure*}[htp]
     \subfloat[][\raggedright Two local Bell pairs are created, swapped to form
     \new{physically} non-local Bell pairs, and then distilled using the
     $Z_{2B}$ protocol~(see Sec.~\ref{sec:Local2To1} for additional details).
     ]{
         \includegraphics[clip,width=.9\columnwidth]{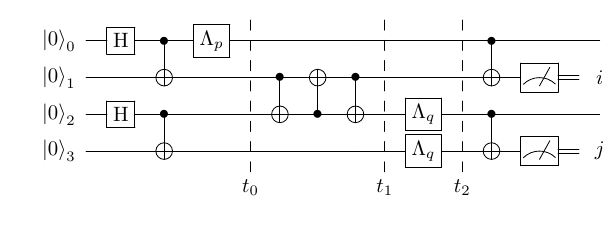} 
     \label{fig:TwoToOne}}
     \qquad
     \subfloat[][\raggedright Three local Bell
     pairs are created, swapped to form \new{physically} non-local Bell pairs, and then
     distilled using the $ZX_{3B}$ distillation protocol~(see Sec.~\ref{sec:Local3To1}
     for additional details).  ]{
         \includegraphics[clip,width=.9\columnwidth]{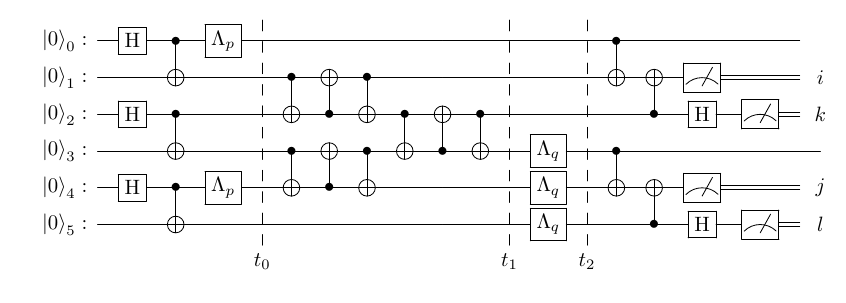} 
     \label{fig:ThreeToOne}}
 \caption{Circuits for entanglement distillation with \new{depolarizing}
    channels $\Lm_p$ and $\Lm_q$ inserted at various stages.}
     \label{fig:circsDist}
\end{figure*}

\begin{figure*}[htp]
    \centering
     \subfloat[][\raggedright Bell pairs are initially prepared with equal Bell fidelity,
     $F_1 = F_2$]{
         \includegraphics[clip,width=.9\columnwidth]{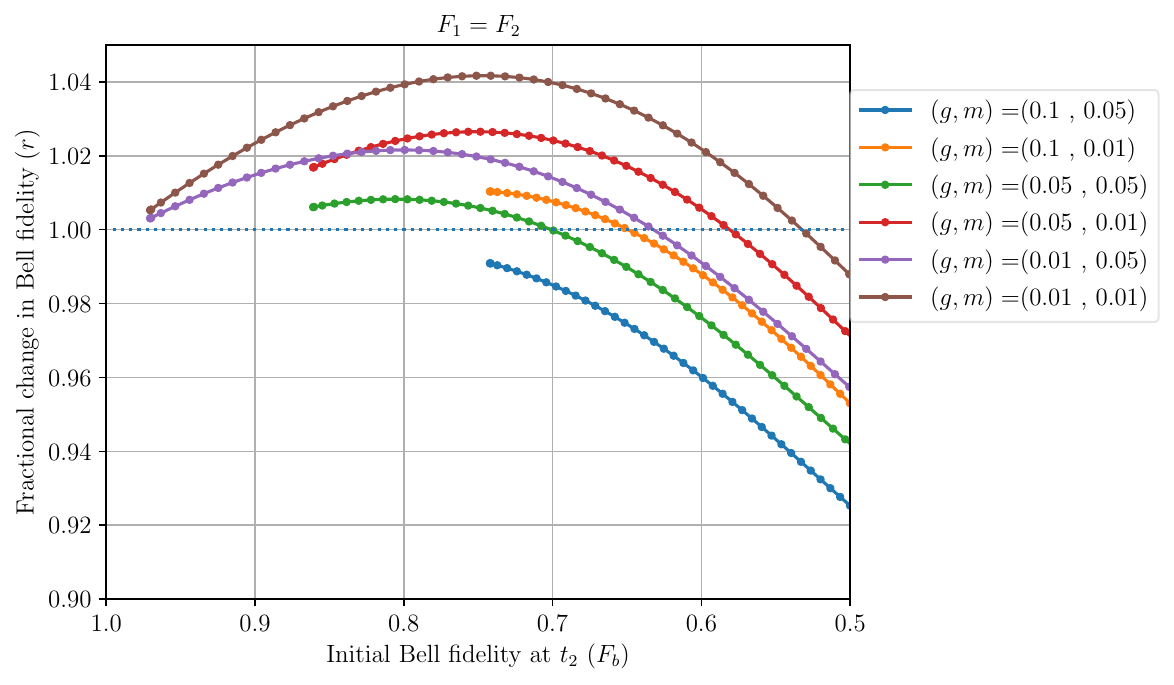}
     \label{fig:2To1Plot1}
     }
     \hfill
     \subfloat[][\raggedright Bell pairs are initially prepared  with unequal
     Bell fidelity, the first Bell pair has $2.5\%$ lower Bell fidelity than
     the second.]{
     \includegraphics[clip,width=.9\columnwidth]{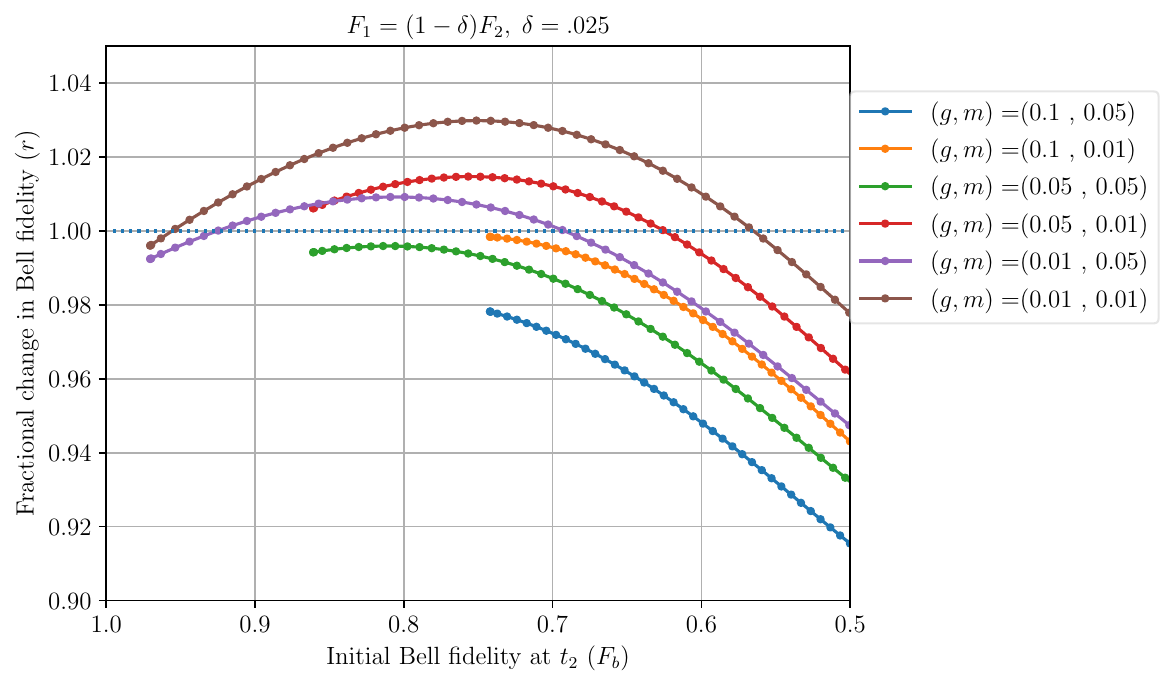}
     \label{fig:2To1Plot2}
     }
     \hfill
     \subfloat[][\raggedright Bell pairs are initially prepared with equal Bell
     fidelity, $F_1 = F_2$]{
         \includegraphics[clip,width=.9\columnwidth]{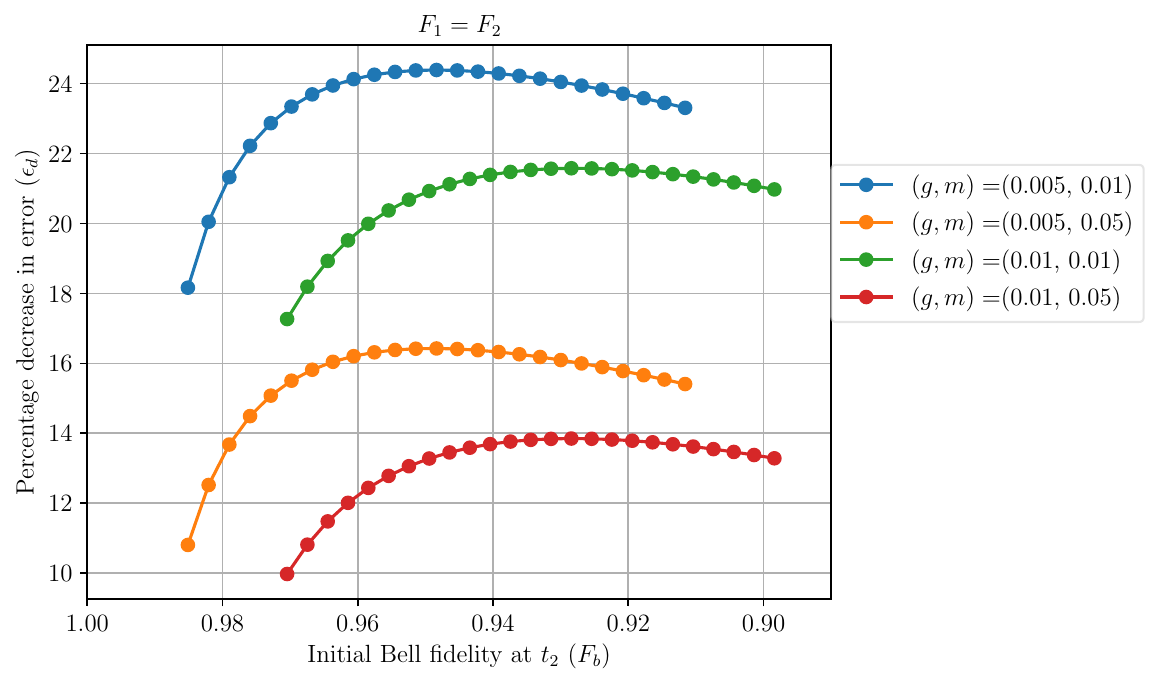}
     \label{fig:2To1Plot3}
     }
     \hfill
     \subfloat[][\raggedright Bell pairs are initially prepared with unequal Bell fidelity,
     the first Bell pair has $2.5\%$ lower Bell fidelity than the second.]{
         \includegraphics[clip,width=.9\columnwidth]{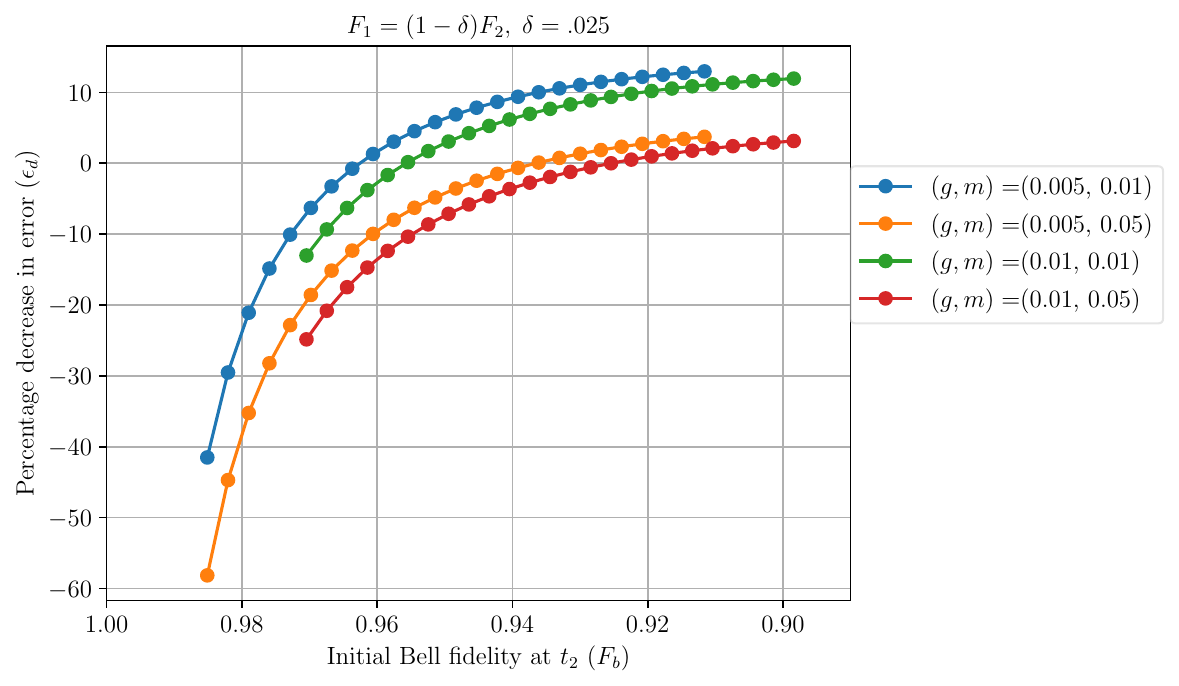}
     \label{fig:2To1Plot4}
     }
     \caption{Results from simulation of recurrence with circuit noise
     described in Fig.~\ref{fig:TwoToOne}. Plots~\eqref{fig:2To1Plot1}
     and~\eqref{fig:2To1Plot2} show fractional change in Bell fidelity,
     $r$~(see eq.~\eqref{eq:ratio}), plotted against initial Bell fidelity,
     $F_b$~(defined below~\eqref{eq:ratio}), for various gate and measurement
     errors, $g$ and $m$, respectively. The region where the plot remain above
     $r=1$ indicates where the noisy distillation circuit is beneficial.
     Plots~\eqref{fig:2To1Plot3} and~\eqref{fig:2To1Plot4} focus on high
     fidelity Bell pairs and show the percentage decrease in Bell infidelity,
     $\ep_d$~(defined in eq.~\eqref{eq:errDec}), plotted against initial
     fidelity $F_b$ for various gate and measurement errors.}
     \label{fig:2To1Plot}
\end{figure*}

We simulate the recurrence protocol ($Z_{2B}$) using the quantum circuit shown
in Fig.~\ref{fig:TwoToOne}. In this circuit all four qubits are initialized to
$\ket{0}$. First the circuit prepares two Bell states~(across qubits $0-1$ and
$2-3$) and then adds noise to the first Bell state~(modelled via a depolarizing
channel $\Lm_p$ acting on the top qubit). This noise helps model asymmetry in
the initial Bell pairs.  In the next part after the first barrier~(dotted line
labelled $t_0$), the circuit swaps one half of each Bell pair creating a
(physically)~non-local Bell pair across qubits $0-2$ and $1-3$. Next, after the
second barrier~(dotted line labelled $t_1$), we apply a waiting error~(modelled
via a one-qubit depolarizing channel $\Lm_q$ acting on qubits 1 and 2 that
constitute one half of each Bell pair). The final part of the circuit, after
the third barrier, carries out a distillation protocol described in
Fig.~\ref{fig:distillBasicZ2B}. All CNOT gates and measurements carry error,
described by channels $\NC_g$ and $\DC_m$, respectively, as discussed in
beginning of this section.

Variation in the performance of distillation with gate error, $g$, measurement
error, $m$, and asymmetry between the initial Bell pairs is plotted in
Fig.~\ref{fig:2To1Plot}. In this figure, the ratio 
\begin{equation}
    r = F_{a}/F_b
    \label{eq:ratio}
\end{equation}
is plotted against $F_b$, where $F_a$ is the Bell fidelity of the distilled
Bell pair and $F_b$ is the maximum of the Bell fidelities among the \new{physically} non-local
Bell pairs just prior to being distilled. The Bell fidelities $F_1$ and $F_2$
are of the Bell pairs just prior to the first barrier in
Fig.~\ref{fig:TwoToOne}

In the first plot, Fig~\ref{fig:2To1Plot1}, the two Bell pairs have equal Bell
fidelity, i.e., $\Lm_p$ in Fig.~\ref{fig:TwoToOne} is noiseless. For any fixed
gate error, $g \in \{.01, .05, .1\}$, and measurement error, $m \in \{.01,
.05\}$, we increase the parameter $q$ in the waiting error $\Lm_q$ and plot
$r$~\eqref{eq:ratio} as a function of the initial Bell fidelity $F_b$. 
In each of these plots as the waiting error $q$ is increased from zero, the
Bell fidelity prior to distillation, $F_b$, decreases while the ratio $r$ first
increases, reaches a maximum \new{and then decreases.}
Across different plots, if we fix the gate error $g$ but increase the
measurement error $m$ we notice a decrease in $r$. On the other hand, if we fix
the measurement error $m$, increasing the gate error $g$ has two effects.
First, it decreases the initial Bell fidelity $F_b$ at $q=0$~(i.e., the plot
begins at a smaller $F_b$ value) and also decreases the initial $r$ value
corresponding to that initial Bell fidelity. Second, it shrinks the interval of
$F_b$ values over which $r>1$.

In the second plot, Fig~\ref{fig:2To1Plot2}, the two Bell pairs have unequal
Bell fidelity, i.e., $\Lm_p$ in Fig.~\ref{fig:TwoToOne} is noisy with $p \neq
0$. After the action of this noisy identity gate, the first Bell pair~(across
qubits $0$ and $1$) becomes more noisy than the second~(across qubits $2$ and
$3$) and Bell fidelity of the first Bell pair $F_1$ is $2.5\%$ lower than that of
the second, $F_2$.
Salient features of the plot in Fig.~\ref{fig:2To1Plot2} resemble those of
Fig.~\ref{fig:2To1Plot1} discussed above. The key difference is that for fixed
$g$ and $m$ values the value of $r$ in Fig.~\ref{fig:2To1Plot2} are smaller
than the corresponding values in Fig.~\ref{fig:2To1Plot1}. In fact, we only see
improvement in a small region where the gate and measurement errors are $ \leq
1\%$. 

In Fig.~\ref{fig:2To1Plot3} and~\ref{fig:2To1Plot4} we focus on smaller gate
and measurement errors, $g = 1 \times 10^{-2}$ or $5 \times 10^{-3}$ and $m = 5
\times 10^{-2}$ or $1 \times 10^{-2}$. Here the $Y$-axis is the percentage
decrease in error, i.e.,
\begin{equation}
    \ep_d = \frac{F_a - F_b}{1-F_b} \times 100,
    \label{eq:errDec}
\end{equation}
while the $X$-axis is the initial Bell fidelity $F_b$.  We obtain points along
these axes by varying the waiting error $q$ in $\Lm_q$.  Here, the initial Bell
fidelity is typically greater than $0.9$ and we notice a $10 \% - 20 \%$
decrease in error by performing distillation on Bell pairs with equal Bell
fidelity~(see Fig.~\ref{fig:2To1Plot3}).  When the Bell pairs have unequal Bell
fidelity, recurrence does not necessarily decrease the error for very high Bell
fidelity but manages to decreases error as the initial Bell fidelities become
smaller~(see Fig.~\ref{fig:2To1Plot4}). This observation is consistent with
results in Fig.~\ref{fig:xDephDiff} which show that the range of initial Bell
fidelities where distillation improves is narrow when the initial Bell
fidelities are high.

\subsubsection{Three Bell Distillation}
\label{sec:Local3To1}

Next, we simulate the $ZX_{3B}$ protocol using the circuit in
Fig.~\ref{fig:ThreeToOne}. This circuit prepares three Bell pairs, which we
label as $AA$, $BB$, and $CC$, initialized on qubits $0-1$, $2-3$, and $4-5$,
respectively. Next, the circuit adds noise to Bell states $AA$ and $CC$ via
$\Lm_p$. After the first barrier~(labelled $t_0$) a sequence of CNOT gates
re-order the Bell pairs into $ABCABC$, such that one half of each Bell pair is
separated from its other half by two qubits. After the second barrier, the
circuit applies a waiting error~(via channel $\Lm_q$ acting on one half of each
Bell pair), and finally, it carries out the $ZX_{3B}$ distillation
protocol~(see Sec.~\ref{sec:ZZZIXX}).

\begin{figure*}[htp]
    \centering
     \subfloat[][\raggedright Bell pairs are initially prepared with equal Bell fidelity, $F_1 = F_3 = F_2$]{
         \includegraphics[clip,width=.9\columnwidth]{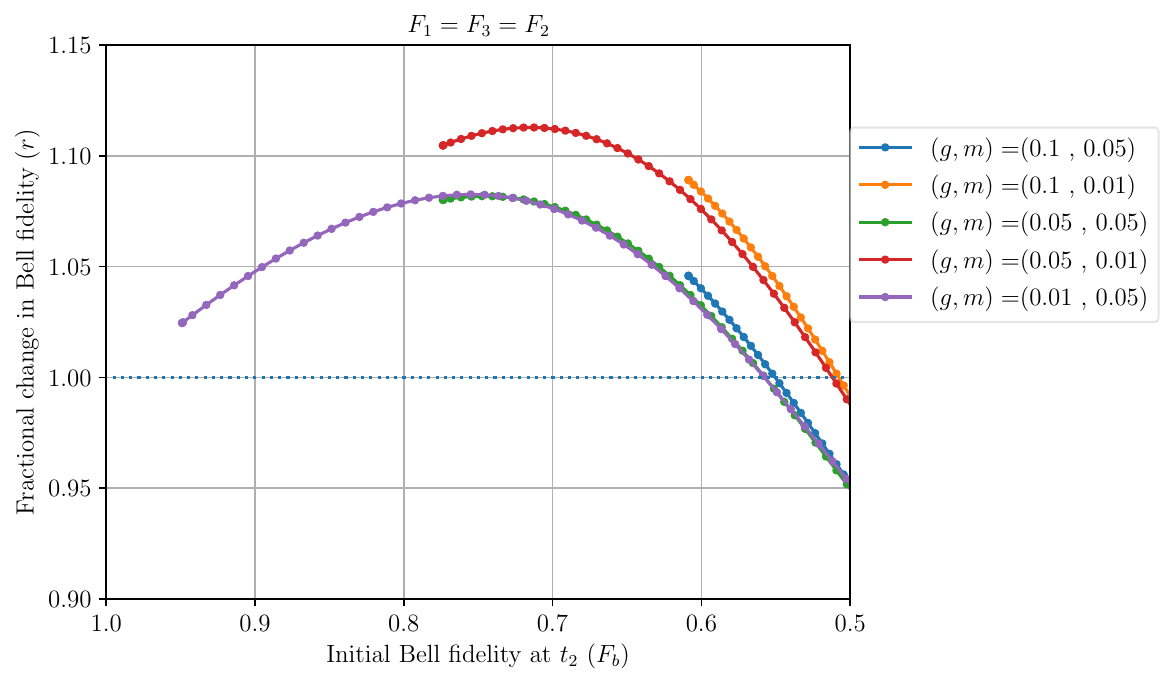}
     \label{fig:3To1Plot1}
     }
     \hfill
     \subfloat[][\raggedright Bell pairs are initially prepared  with unequal Bell fidelity,
     the first and third Bell pair has $2.5\%$ lower Bell fidelity than the
     second]{
     \includegraphics[clip,width=.9\columnwidth]{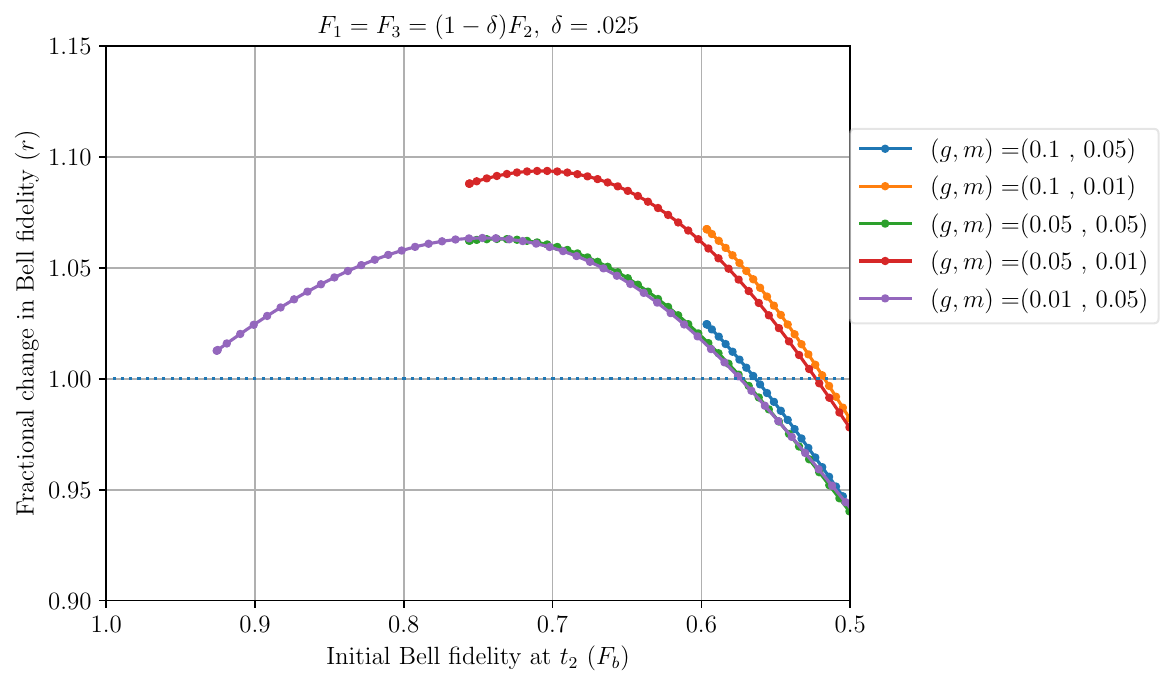}
     \label{fig:3To1Plot2}
     }
     \hfill
     \subfloat[][\raggedright Bell pairs are initially prepared with equal Bell fidelity, $F_1 = F_3 = F_2$]{
         \includegraphics[clip,width=.9\columnwidth]{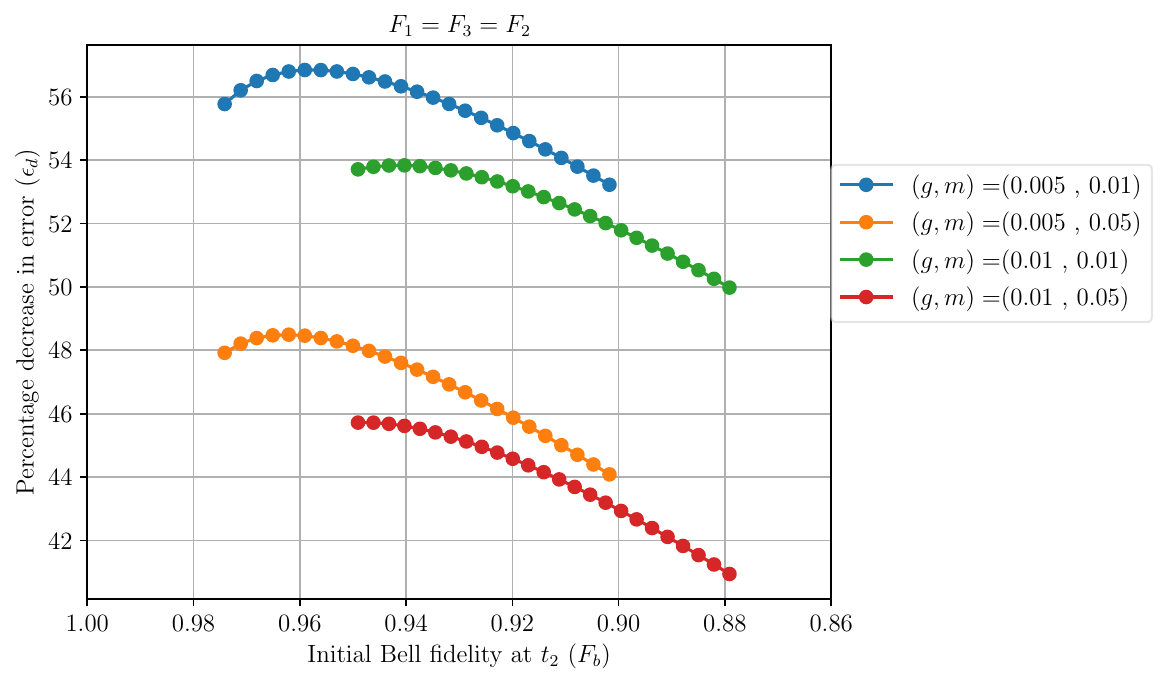}
     \label{fig:3To1Plot3}
     }
     \hfill
     \subfloat[][\raggedright Bell pairs are initially prepared with unequal Bell fidelity,
     the first and third Bell pairs have $2.5\%$ lower Bell fidelity than the second.]{
         \includegraphics[clip,width=.9\columnwidth]{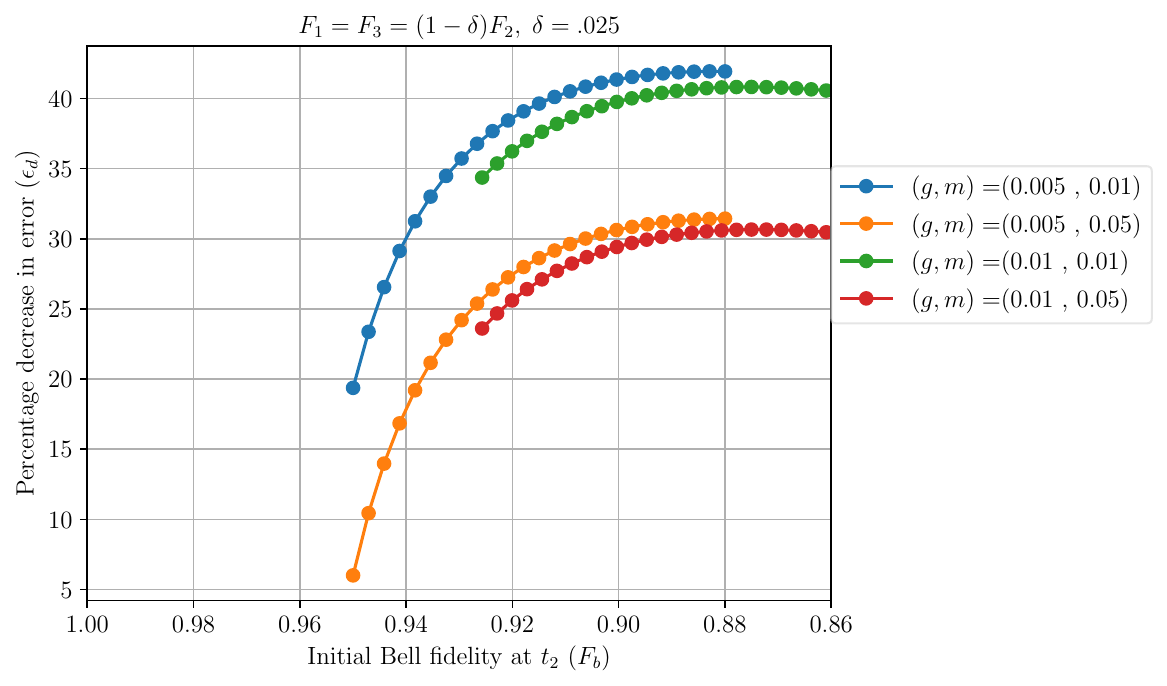}
     \label{fig:3To1Plot4}
     }
     \caption{Results from simulation of $ZX_{3B}$ distillation with circuit
     noise given in Fig.~\ref{fig:ThreeToOne}.  Plots~\eqref{fig:3To1Plot1}
     and~\eqref{fig:3To1Plot2} show fractional change in Bell fidelity,
     $r$\new{~(see eq.~\eqref{eq:ratio})}, plotted against initial Bell fidelity,
     $F_b$\new{~(defined below~\eqref{eq:ratio})}, for various gate and
     measurement errors. The region of the plot above $r=1$ indicates where the
     noisy distillation circuit is beneficial.  Plots~\eqref{fig:3To1Plot3}
     and~\eqref{fig:3To1Plot4} focus on low noise Bell pairs and show the
     percentage decrease in Bell infidelity, \new{$\ep_d$~(defined in
     eq.~\eqref{eq:errDec})}, plotted against initial fidelity $F_b$ for
     various gate and measurement errors.}
     \label{fig:3To1Plot}
\end{figure*}

The first plot in Fig.~\ref{fig:3To1Plot1} describes the variation in the ratio
$r$ as a function of the $F_b$ for fixed gate and measurement errors. All three
Bell pairs have the same initial Bell fidelity. The variation in $r$ and $F_b$
with the gate error $g$ and measurement error $m$ are similar to those in
Fig.~\ref{fig:2To1Plot1}, described below eq.~\eqref{eq:ratio}.
For fixed $g$ and $m$, the ratio $r$ in Fig.~\ref{fig:3To1Plot1} is typically
higher than those in Fig.~\ref{fig:2To1Plot1} when $r>1$. In addition, the
value of $F_b$ at which $r$ shifts from a value greater than one to a value
less than one is typically lower in Fig.~\ref{fig:3To1Plot1} compared to
Fig.~\ref{fig:2To1Plot1}, i.e., the parameter region and amount by which
distillation provides an improvement seem to be typically larger in
Fig.~\ref{fig:3To1Plot1} compared to Fig.~\ref{fig:2To1Plot1}.

In Fig.~\ref{fig:3To1Plot2} two of the Bell pairs, the first $AA$ pair across
qubits $0-1$ and third $CC$ across qubits $4-5$, have lower Bell fidelity, than
the second pair $BB$ across $2-3$, i.e.  $F_1 = F_3 < F_2$.  The noise
parameter $p$ is such that the initial Bell fidelity $F_1 = F_3  = .975 F_2$.
As a result distillation protocol here can be seen as an attempt to improve the
fidelity of one Bell pair using two Bell pairs with lower fidelity.
Variation in improvement $r$ and $F_b$ with the gate error $g$ and measurement
error $m$ is similar to those in Fig.~\ref{fig:2To1Plot2}, described below
eq.~\eqref{eq:ratio}. For fixed $g$ and $m$, the ratio $r$ in
Fig.~\ref{fig:3To1Plot1} can be higher than those in
Fig.~\ref{fig:2To1Plot2}, however this need not be the case in general, even
when $r>1$

The focus in Fig.~\ref{fig:3To1Plot3} and~\ref{fig:3To1Plot4} is on smaller
gate and measurement errors, $g = 1 \times 10^{-2}$ or $5 \times 10^{-3}$ and
$m = 5 \times 10^{-2}$ or $1 \times 10^{-2}$. As the error parameter $q$ is
varied, we plot the percentage decrease in error against the initial Bell
fidelity $F_b$. When the Bell pairs being distilled have equal fidelity
initially we find a $40 \% - 50 \%$ decrease in error, as shown in
Fig.~\ref{fig:3To1Plot3}. The decrease in error is lower when the when Bell
pairs have unequal Bell fidelities initially, as shown in
Fig.~\ref{fig:3To1Plot4}; it is possible that error increases if the initial
Bell fidelities are made even more unequal.

\subsection{Global Depolarizing noise}
\label{sec:globalDepol}
In this section, we apply circuit noise to Bell pairs and degrade them under a
global depolarizing noise channel $\NC_{\lm}$~(see Fig.~\ref{fig:circsGl} in
App.~\ref{app:circDst}); this global noise is consistent with Clifford
twirling~\cite{EmersonEA05, DankertEA09}, which we will do experimentally in
Sec.~\ref{sec:glDplExp}.

\subsubsection{Two Bell Distillation}
\label{sec:Global2To1}

\begin{figure*}[htp]
    \centering
     \subfloat[][\new{Bell pairs are prepared
     with equal Bell Fidelity.}]{
         \includegraphics[clip,width=.9\columnwidth]{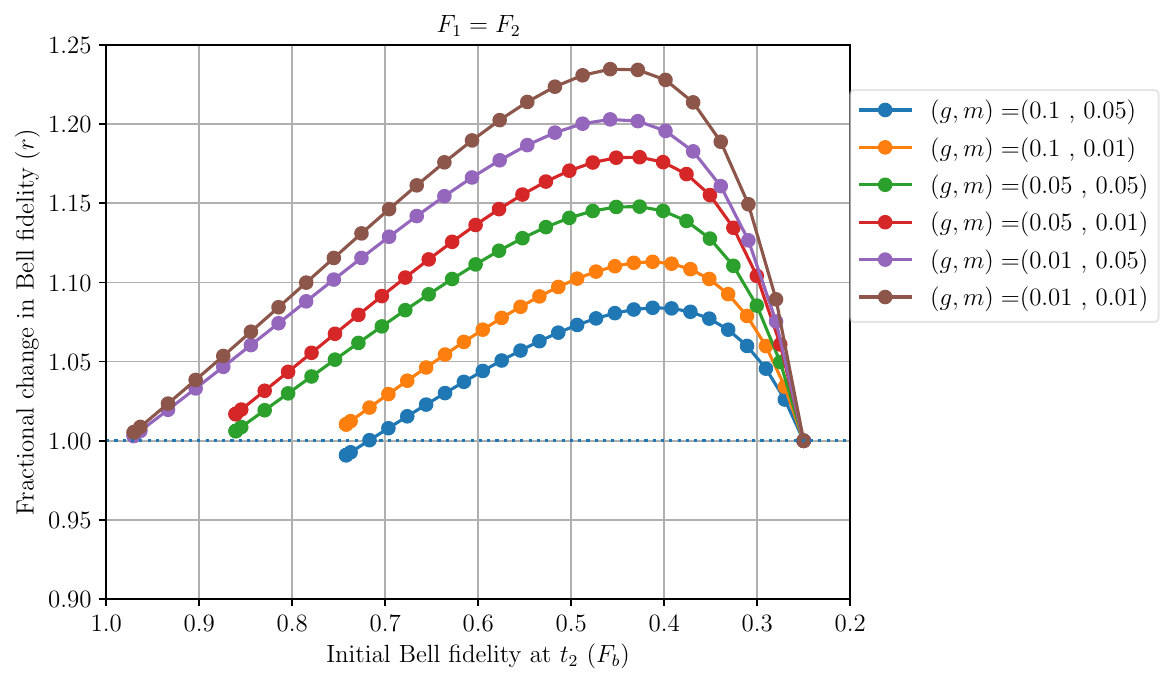}
     \label{fig:2To1PlotGl1}
     }
     \qquad
     \subfloat[][\new{Bell pairs are prepared
     with unequal Bell fidelity, the second Bell pair has $9 \%$ lower Bell
     fidelity than the second.}]{
         \includegraphics[clip,width=.9\columnwidth]{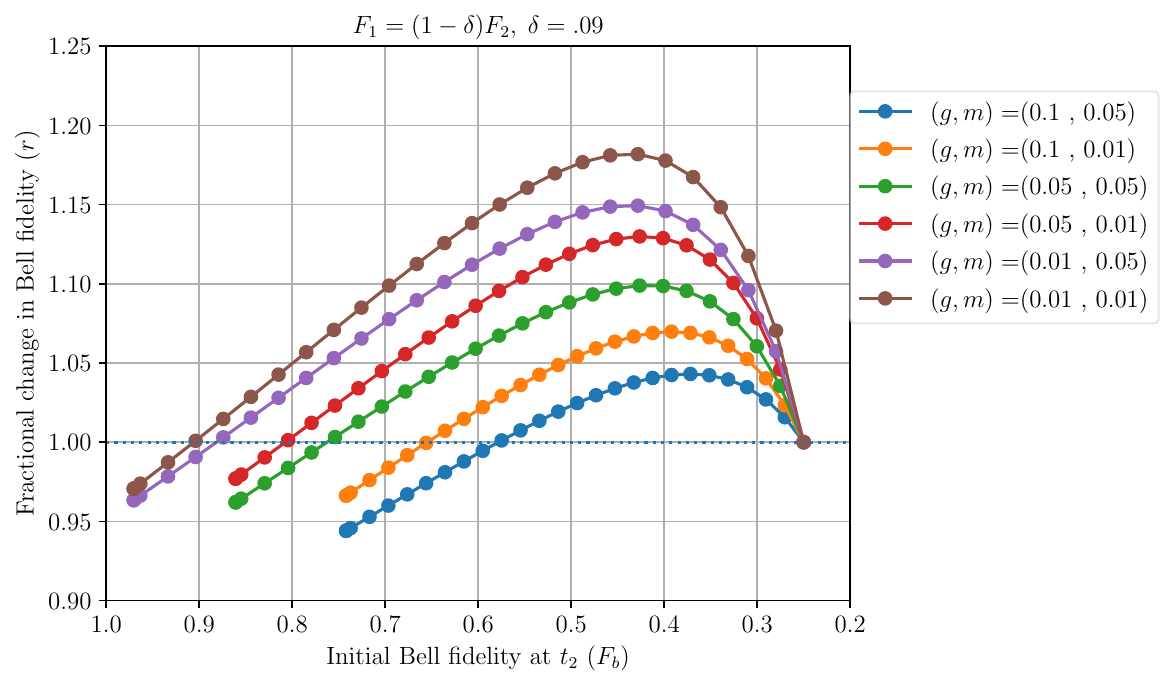}
     \label{fig:2To1PlotGl2}
     }
     \caption{\new{Results from simulation of recurrence $Z_{2B}$ with global
     depolarizing noise~(for circuit see Fig.~\ref{fig:TwoToOneGl} in
     App.~\ref{app:circDst}). Plots~\eqref{fig:2To1PlotGl1}
     and~\eqref{fig:2To1PlotGl2} show fractional change in Bell fidelity,
     $r$ defined in eq.~\eqref{eq:ratio}, plotted against initial Bell
     fidelity, $F_b$~(defined below eq.~\eqref{eq:ratio}), for various gate and
     measurement errors. The region of the plot above $r=1$ indicates where the
     noisy distillation circuit is beneficial.}}

     \label{fig:2To1PlotGlInf}
\end{figure*}

We numerically study the performance of the recurrence distillation protocol
($Z_{2B}$) after the Bell pairs are degraded by a global depolarizing channel.
The distillation circuit under study modifies the one in
Fig.~\ref{fig:TwoToOne} by replacing local with global depolarizing noise
during waiting~(circuit available as Fig.~\ref{fig:TwoToOneGl} in
App.~\ref{app:circDst}).
In the first plot, see Fig.~\ref{fig:2To1PlotGl1}, the Bell pairs are created
with equal Bell fidelity, $F_1 = F_2$. As the global depolarizing noise
parameter $\lm$ is increased, the fidelity of Bell pairs prior to distillation
decreases. For any fixed gate and measurement error, as $\lm$ is increased the
ratio $r$, plotted on the y-axis, first increases reaches a maximum and then
decreases to one. This ratio can be as high as 1.2, and it remains above one
for a wider range of gate and measurement error values.  This represents broad
improvement from distillation. The improvements shrink as the gate and
measurement error are increased. In Fig.~\ref{fig:2To1PlotGl2} the Bell pairs
have unequal Bell fidelity prior to distillation. This asymmetry is created by
noise channel $\Lm_p$ on qubit labelled 0 in Fig.~\ref{fig:TwoToOneGl}. A key
qualitative effect of this asymmetry is in the ratio $r$. This ratio remains
below one for higher initial Bell fidelities, thus shrinking the range of gate
and measurement error values over which distillation presents an improvement.
The improvement is also smaller in comparison to those reported in
Fig.~\ref{fig:2To1PlotGl1}.

\subsubsection{Three Bell Distillation}
\label{sec:Global3To1}

Here we continue from the previous section, but indicate results from the
double selection ($ZX_{3B}$) protocol. The effect of global depolarizing noise
on the $ZX_{3B}$ distillation protocol is similar to that of recurrence shown
in Fig.~\ref{fig:2To1PlotGlInf} and described in Sec.\ref{sec:Global2To1}.
However the value of $r$ is typically higher than those in the corresponding
plots on Fig.~\ref{fig:2To1PlotGlInf}~(see App.~\ref{app:dataDist} for plots on
Fig.~\ref{fig:3To1PlotGlobal}).

\section{Experimental Results and Device Noise}
\label{sec:deviceNoise}

Given the simulations of the previous sections, we now want to consider
distillation protocols on a real device. We run experiments on a 127 qubit
device with fixed frequency qubits and fixed coupling, an IBM \texttt{Eagle} device
\texttt{ibm\_kyiv}. For the purposes of these experiments we consider smaller sections
of the device with linear connectivity and run each experiment in parallel
across different sections of the device in order to obtain more statistics on
noise. For more device details see Appendix~\ref{ap:device}. 

For each circuit we estimate the fidelity of the Bell state using direct
fidelity estimation by measuring three circuits~(see App.~\ref{ap:DirBell} for
more). However, this scheme does have measurement error. In contrast
to~\cite{YanZhongEA22}, here we choose to not correct for measurement errors
due to issues that can cause potentially non-physical and or unexpectedly high
fidelities (see, e.g., discussion in Ref.~\cite{Gupta2024}). Our experimental
results thus provide a good lower bound on the Bell fidelities.

\subsection{Two Bell distillation experiment with global depolarizing noise}
\label{sec:glDplExp}

Following from the discussion on Sec.~\ref{sec:globalDepol}, first, we
experimentally study the recurrence protocol under global depolarizing noise.  
This global depolarizing channel $\NC_{\lm}$ is implemented using layers of
two-qubit Clifford circuits.
The first half of such a circuit is composed of multiple applications of a
length-$k$ random sequence of two-qubit  Clifford gates and the second half is
the inverse of the first half.  This inverse is simply the mirror image of the
first half and these layers are called mirror Clifford
layers~\cite{ProctorEA22}~(see Fig.~\ref{fig:Layer} in App.~\ref{app:circDst}
for an example of such a circuit with two layers in the first half).

\begin{figure*}[htp]
    \centering
     \subfloat[][Acceptance ratio]{\includegraphics[clip,width=.9\columnwidth]{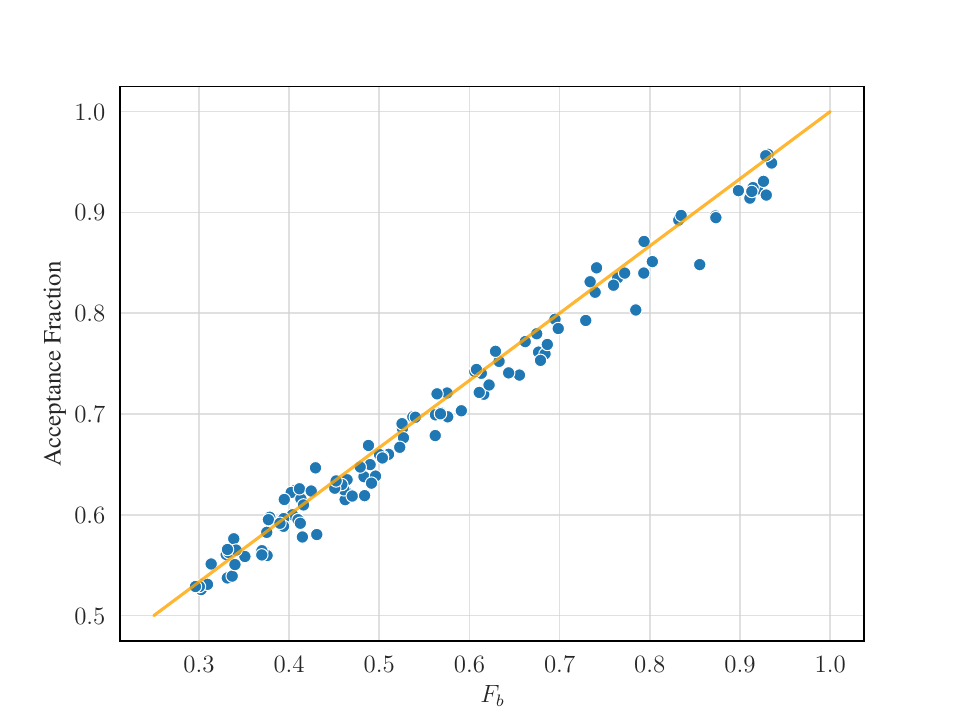}
     \label{fig:AccRatioGlobal}
     }
     \qquad
     \subfloat[][Fractional change in Bell fidelity with distillation]{\includegraphics[clip,width=.9\columnwidth]{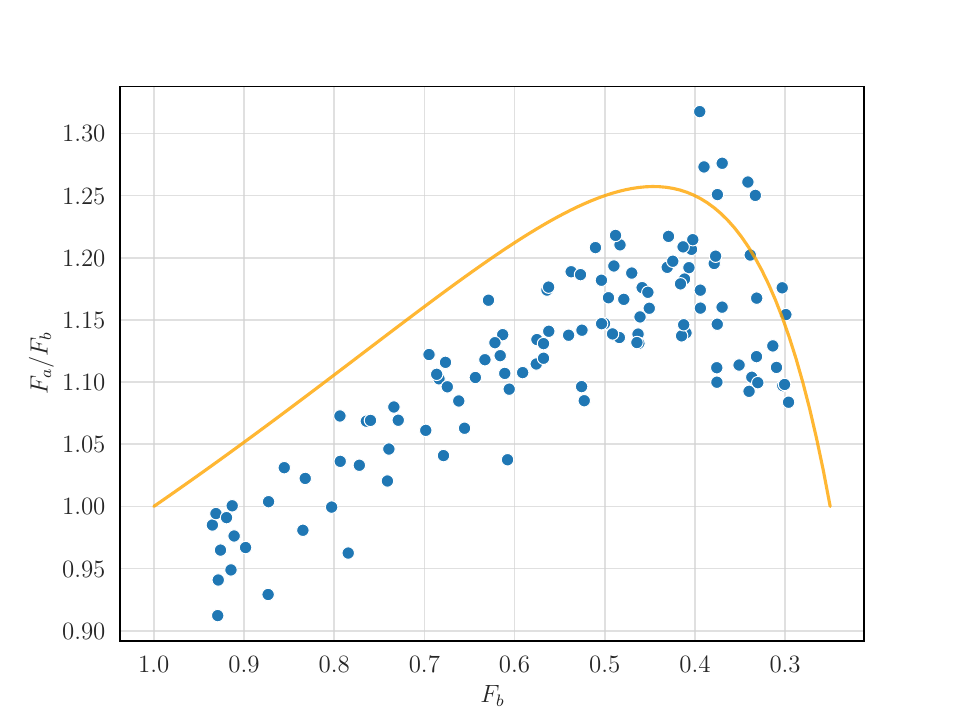}
     \label{fig:FidRatioGlobal}
     }
     \caption{Experimental results (blue) on \texttt{ibm\_kyiv} for the $Z_{2B}$ recurrence
     distillation protocol with global depolarizing noise. Orange curves are
     theory assuming perfect distillation.}
     \label{fig:GlobalExp}
\end{figure*}

These measurements are carried out on a set of four adjacent qubits on a
device.  On 13 such sets we carry out the experiment outlined above for
different fixed $k \in \{0,2,4,6,8,10,12\}$. For each $k$ we obtain an average
of 13 sets of the before and after distillation Bell fidelities along with the
acceptance probability for distillation. These are plotted in
Fig.~\ref{fig:GlobalExp}.
In Fig.~\ref{fig:AccRatioGlobal} we plot the acceptance \new{fraction} 
as a function of
the initial Bell fidelity. The dots represent points from the experiment while
the straight line is obtained from numerics. In Fig.~\ref{fig:FidRatioGlobal}
we plot the ratio $r$ as a function of the initial Bell fidelity prior to
global depolarizing noise. The blue dots represent data while the orange curve
represents the theoretically expected curve. This theoretical curve from
eq.~\eqref{eq:RecurGlobal} assumes noiseless gates and measurement. There is
fairly good agreement between the theory curves and the experiment data,
particularly considering the x-axis is subject to systematic error due to
measurement error and the theory calculation assumes the distillation circuit
is perfect. This demonstrates that under some conditions on the device we can
see an improvement due to distillation, however the improvement arises from a
systematic procedure to degrade the Bell pairs via global depolarizing noise.
Furthermore, high fidelity Bell pairs do not improve in this experiment.

\subsection{Two Bell distillation experiment with idle noise}
\label{sec:Device2To1}

\begin{figure}[htp] 
    \includegraphics[scale=.5]{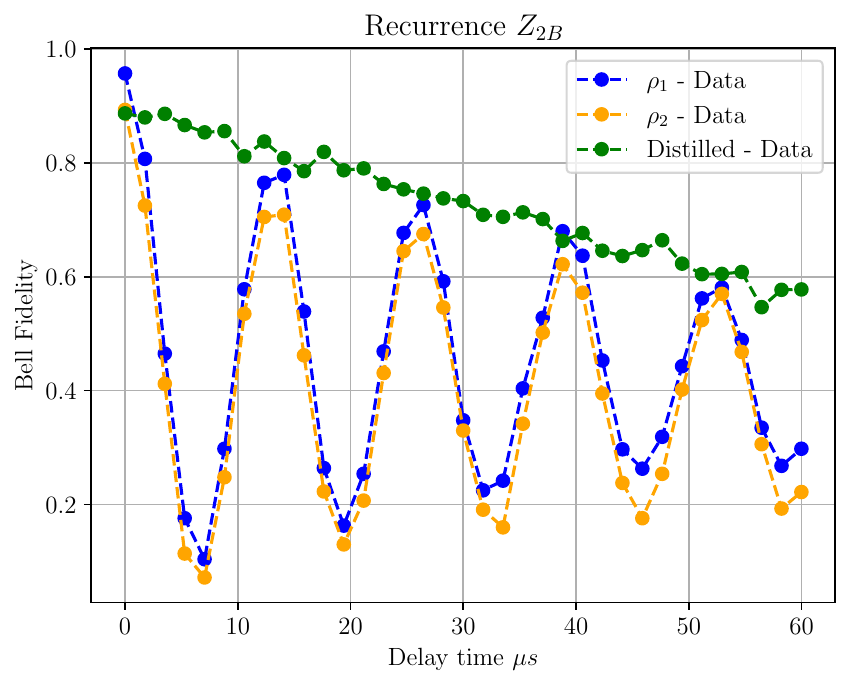}
    \caption{Experimental data from qubits (0,1,2,3) of \texttt{ibm\_kyiv} (points)
    showing increase in the Bell fidelity under $Z_{2B}$ distillation of qubits
    undergoing coherent $ZZ$ errors.}
    \label{fig:cohErr}
\end{figure}

Here we consider a more natural experiment, create a Bell pair and wait a time
$t$ for the Bell fidelity to decay under device $T_1/T_2$ noise and then
measure (note we apply a standard echo sequence to remove low frequency noise).
These $T_1/T_2$ noise channels are not simply Pauli channels and they are not
an effective error obtained from averaging over random iterations, as we did in
the previous section. However, an effective $T_1/T_2$ idling error is not
obtained by simply waiting on an actual device; due to $ZZ$ interactions
between neighboring pairs the Bell state fidelity can strongly degrade. We show
a simple example of this in Fig.\eqref{fig:cohErr}. Here we can get tremendous
improvement in the Bell fidelity, but this is because we allowed the Bell
fidelity to decrease via a coherent error term; the purity of the state is
unchanged. One has to be careful of similar situations where distillation
provides an apparent benefit, but where improvement could more easily be
achieved without post-selection. In this example, the $ZZ$ coherent error term
can be also be canceled by applying staggered dynamical
decoupling~(DD)~\cite{ZeyuanEA23, LiranEA24, SeifEA24}~(see
App.~\ref{app:circDst}, Fig.~\ref{fig:DeviceExp}). With staggered DD, the
circuit is now well described by $T_1$ and $T_2$ errors.

\begin{figure*}[htp]
    \centering
     \subfloat[\raggedright Bell fidelity of the \new{physically} non-local Bell pairs prior to 
        and after recurrence $Z_{2B}$. ]
        {
         \includegraphics[clip,width=.9\columnwidth]{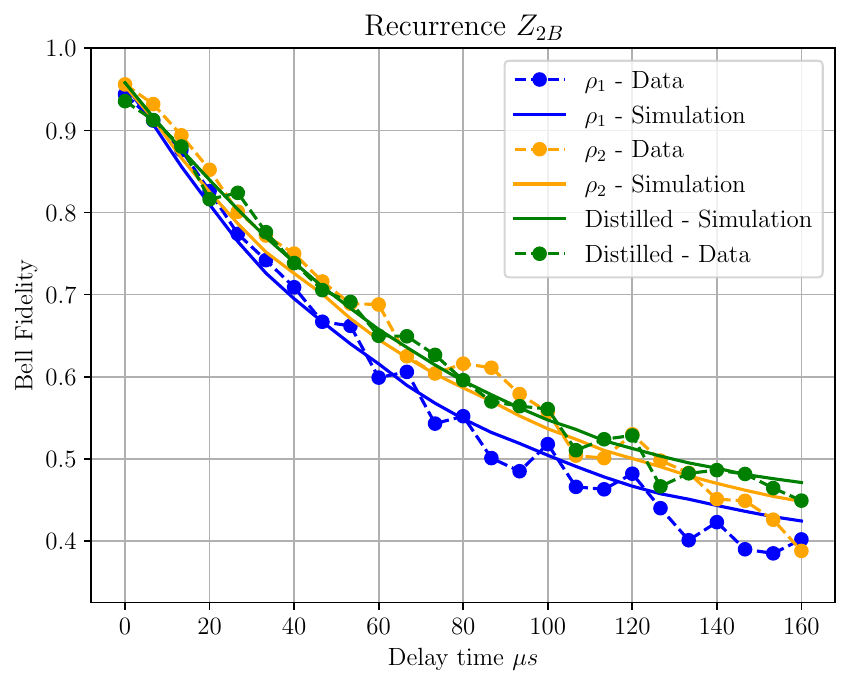}
     \label{fig:XCheckBellAll}
     }
        \hfill
     \subfloat[][\raggedright Bell fidelity of the \new{physically} non-local Bell pairs prior to 
        and after recurrence $X_{2B}$.]{
         \includegraphics[clip,width=.9\columnwidth]{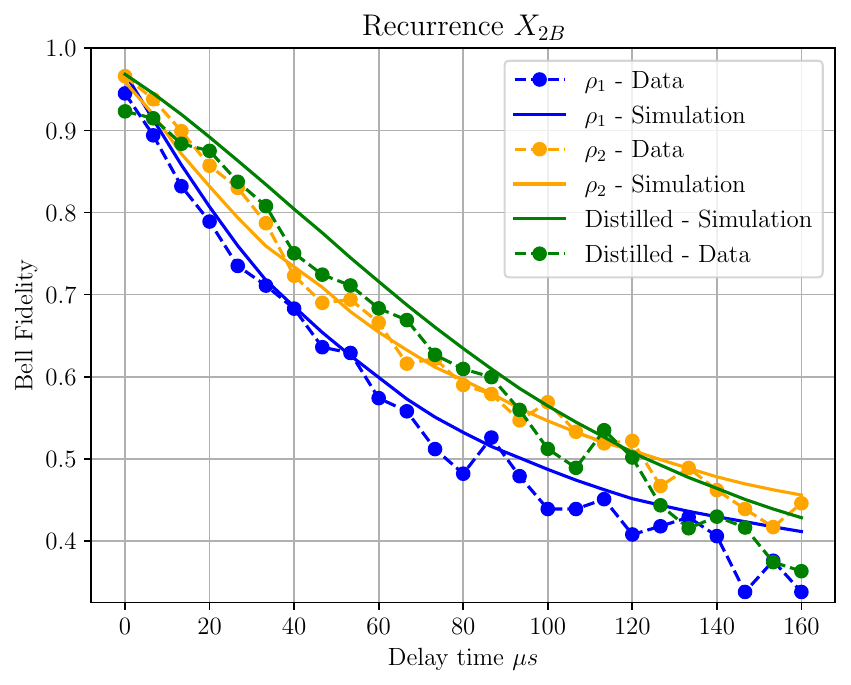}
     \label{fig:ZCheckBellAll}
     }
     \hfill
        \subfloat[][Fractional change in Bell fidelity with recurrence $Z_{2B}$.]{
         \includegraphics[clip,width=.9\columnwidth]{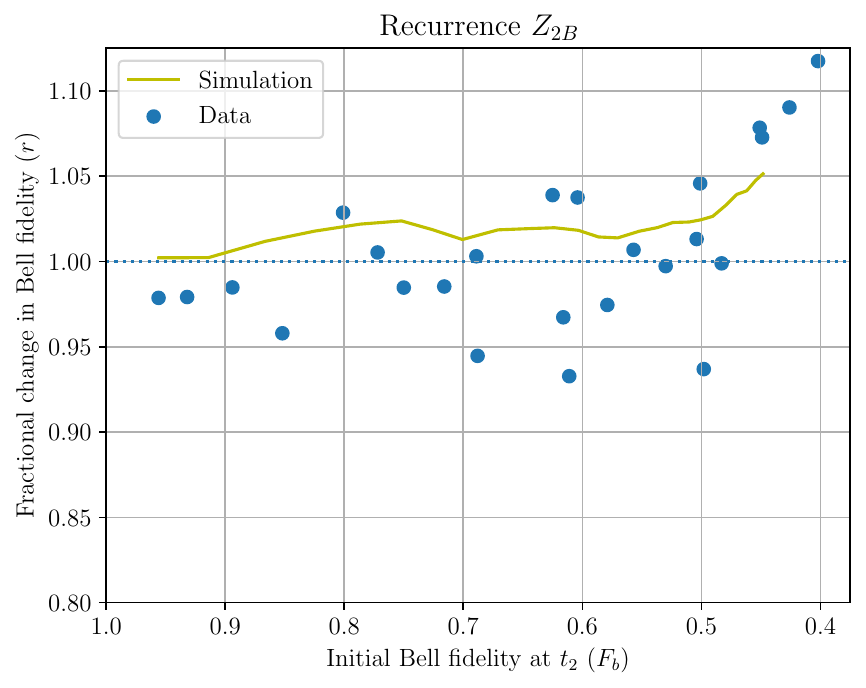}
        \label{fig:XCheckRatio}
        }
        \hfill
        \subfloat[][Fractional change in Bell fidelity with recurrence $X_{2B}$.]{
         \includegraphics[clip,width=.9\columnwidth]{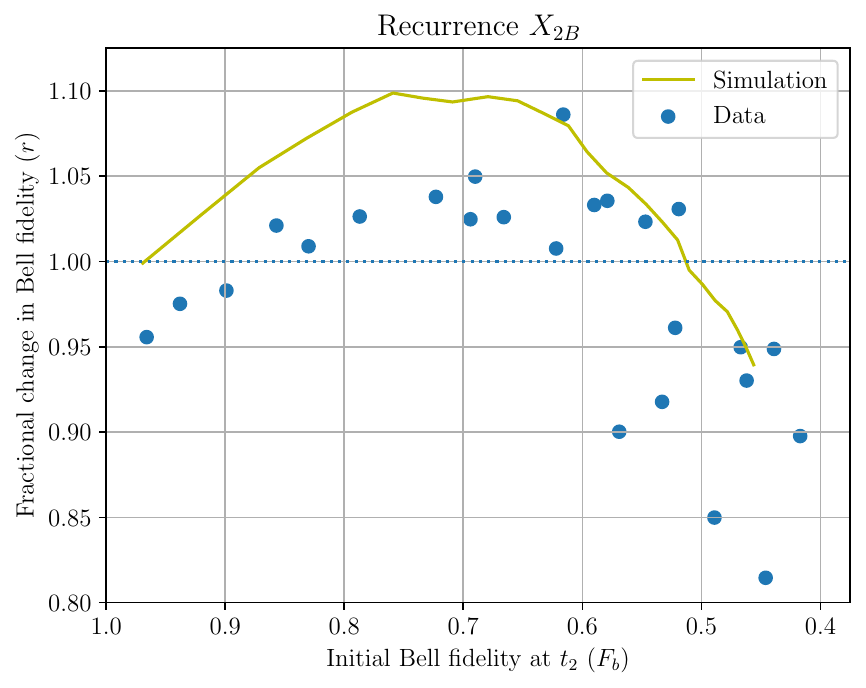}
        \label{fig:ZCheckRatio}
        }
     \caption{Data and simulation for recurrence $Z_{2B}$ and $X_{2B}$ on
     qubits (0,1,2,3) of \texttt{ibm\_kyiv}. \new{In Fig.~\ref{fig:XCheckBellAll}
     and~\ref{fig:ZCheckBellAll} the Bell fidelity for various states is plotted
     against the delay time.  States $\rho_1$ and $\rho_2$ label the different
     physically non-local Bell pairs being distilled while the state obtained
     after distillation appears separately.} Solid lines are
     simulations assuming reported noise parameters as described in
     Table.~\ref{table:2To1DevPar_Z} and~\ref{table:2To1DevPar_X} of
     App.~\ref{ap:device} \new{while dashed-dotted lines represent experimental data.}
     \new{Figs.~\ref{fig:XCheckRatio} and~\ref{fig:ZCheckRatio} plot the
     fractional change, $r$ defined in eq.~\eqref{eq:ratio}, against the initial
     Bell fidelity, $F_b$ defined below eq.~\eqref{eq:ratio}.} Additional plots
     in App.~\ref{app:dataDist}, Fig.~\ref{figApp:2To1ExpSim}
     and~\ref{figApp:2To1ExpSimZCheck}}
     \label{fig:2To1ExpSim}
\end{figure*}

Once we apply staggered DD, typical data is shown in Fig.~\ref{fig:2To1ExpSim}
with dotted lines. We see some natural features of the experiments.  One, the
starting Bell fidelity is lower than one due to the creation and swap
operations~(for instance see the Bell fidelity in Figs.~\ref{fig:XCheckBellAll}
and~\ref{fig:ZCheckBellAll} at $t=0$ is less than one).
Second, there is a natural asymmetry in the initial fidelity due to the
variability of noise parameters between qubits~(for instance the Bell fidelity
of the two bell pairs in Fig.~\ref{fig:ZCheckBellAll} are unequal). 
Third, depending on the distillation protocol used, $Z_{2B}$ or $X_{2B}$, one
may or may not see broad improvement in Bell fidelity~(see
Fig.~\ref{fig:ZCheckRatio} with improvement and~\ref{fig:XCheckRatio} without);
however we do not yet see broad across the device.  To understand these
features better we look to build a more involved device noise model.

\subsection{Device Model and Numerics}
\label{sec:devmodel}

To build a more involved device model we must include $T_1/T_2$
noise~\cite{AliferisBritoEA09, SiddhuAbdelhadiEA24}~(see
App.~\ref{ap:T1T2chan} for a mathematical description) during the wait
periods. These include when the swap is occurring, when the Bell states are
idle and when the measurements are occurring.  Furthermore, due to the effect
described in Ref.~\cite{JurcevicGovia22} it's not enough to use the bare
$T_1/T_2$ even though we performed staggered echo sequences.  We must still
include the $ZZ$ terms and simulate the echo sequence to get the proper decay
of the Bell states during the wait. We continue to add in the gate and
measurement noise as described in Sec.~\ref{sec:localDepol}. We pull the $ZZ$,
$T_1$, $T_2$, gate times and measurement error from the backend of the
device~(see Tables~\ref{table:2To1DevPar_Z} and~\ref{table:2To1DevPar_X} in
App.~\ref{ap:device}).

For recurrence $Z_{2B}$ the simulations are given by the solid lines in
Fig.~\ref{fig:XCheckBellAll} and we notice fairly good agreement between the
experimental and numerical simulations.  Confirming the general trend of the
data is consistent with the calibration data.  However, the figure does not
indicate consistent improvement in Bell fidelity with distillation.  To get
improvements we notice~(see Table~\ref{table:2To1DevPar_Z} in
App.~\ref{ap:device}) that, in general, the dominant source of noise on the
qubits under consideration is $T_2$ noise~(i.e., $Z$ errors) while the $Z_{B2}$
catches $T_1$ noise, i.e., $X$ and amplitude damping type errors.

The $X_{2B}$ distillation protocol catches $Z$ errors. Results from
implementing this protocol in both simulation and experiment is given in
Fig.~\ref{fig:ZCheckBellAll}.  In this figure, there is reasonable agreement
between the experimental data and numerical simulations. In addition, notice
there is general improvement in Bell fidelity, indicating that the likely
dominant source of $T_2$ errors are caught by $X_{2B}$ recurrence.
Improvements are not seen for higher Bell fidelities. This likely occurs due to
both measurement errors and asymmetry in Bell fidelities of the two Bell pairs
being distilled. The former are not removed by distillation while the latter
generally shrink the noise region where distillation shows improvement. In
specific cases simulations indicate that lowering measurement errors to zero
can allow distillation to give improvements even for high fidelity Bell pairs. 

\begin{figure*}[htp]
    \centering
     \subfloat[][\raggedright Bell fidelity of the \new{physically} non-local Bell pairs prior
     to and after distillation on qubits (59,60,61,62,63,64).  Simulations do
     not fit the data well; one possibility is that the coherence varied
     between the distillation run and when the coherence was measured.]{
         \includegraphics[clip,width=.9\columnwidth]{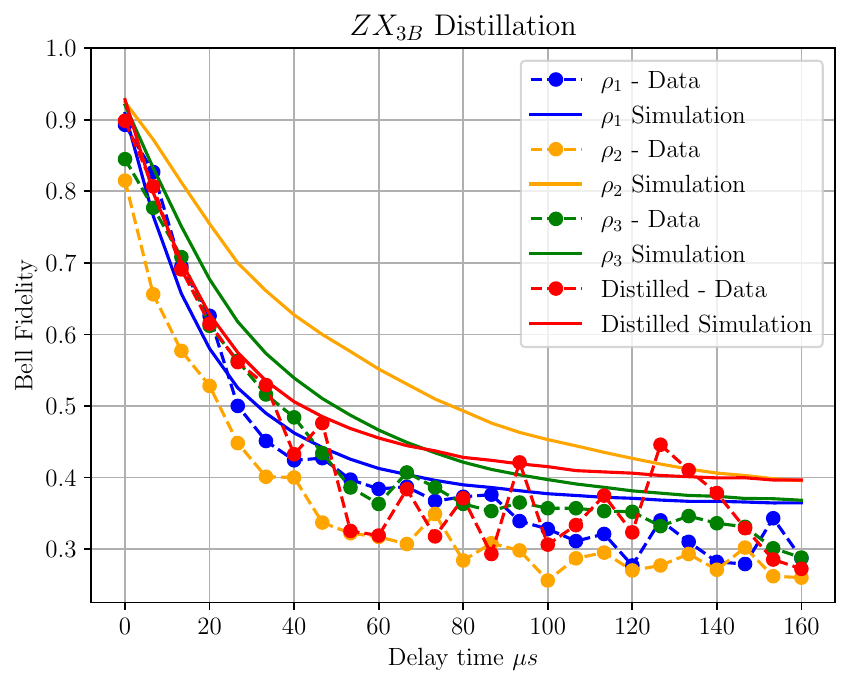}
     \label{fig:3To1Check2BellAll}
     }
        \hfill
     \subfloat[\raggedright Bell fidelity of the \new{physically} non-local Bell pairs prior to 
        and after distillation on qubits (3,4,5,6,7,8). Simulations fits the data
         well.]
        {
         \includegraphics[clip,width=.9\columnwidth]{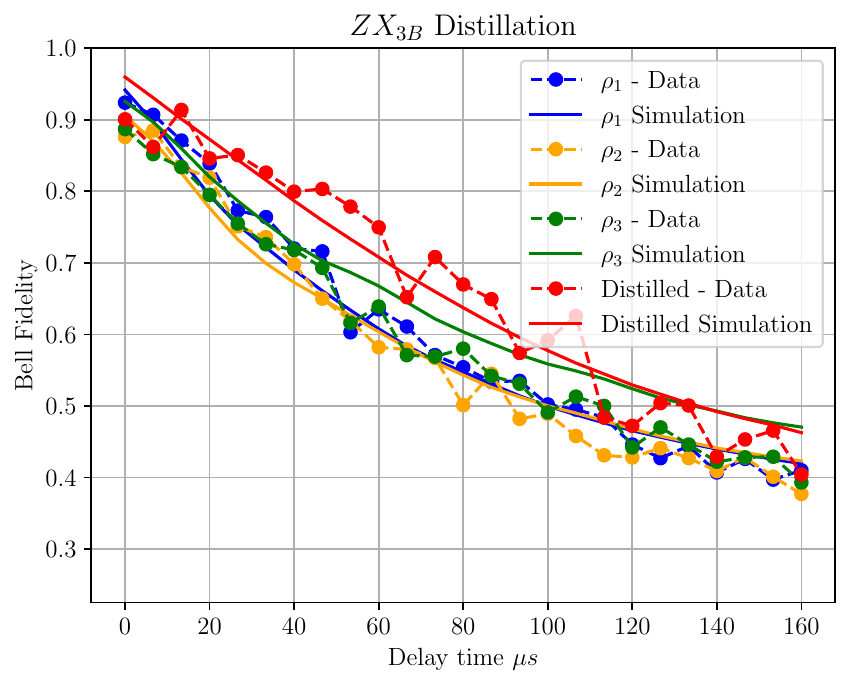}
     \label{fig:3To1CheckBellAll}
     }
     \hfill
        \subfloat[][Fractional change in Bell fidelity on qubits (59,60,61,62,63,64)]{
         \includegraphics[clip,width=.9\columnwidth]{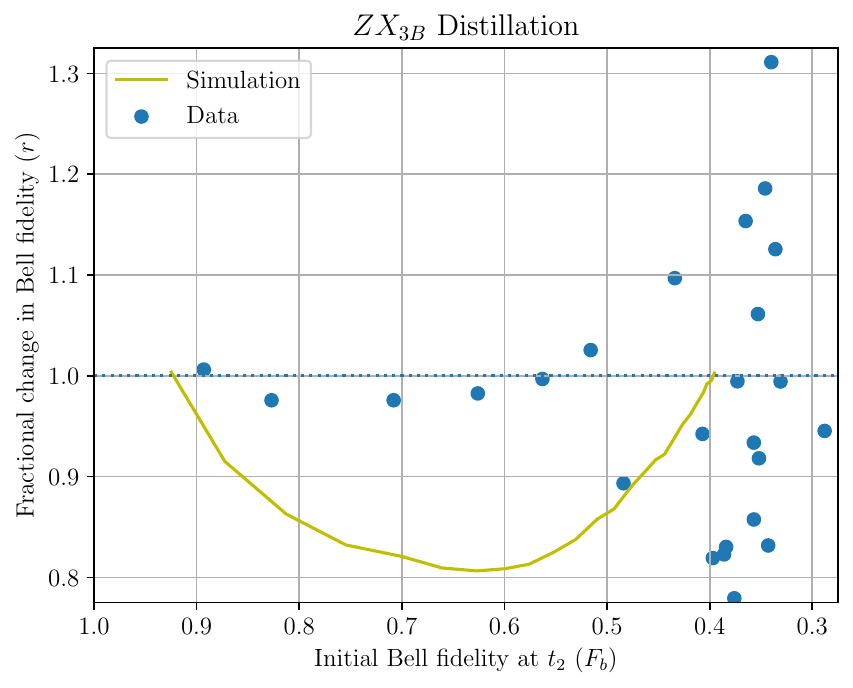}
        \label{fig:3To1Check2Ratio}
        }
        \hfill
        \subfloat[][Fractional change in Bell fidelity on qubits (3,4,5,6,7,8)]{
         \includegraphics[clip,width=.9\columnwidth]{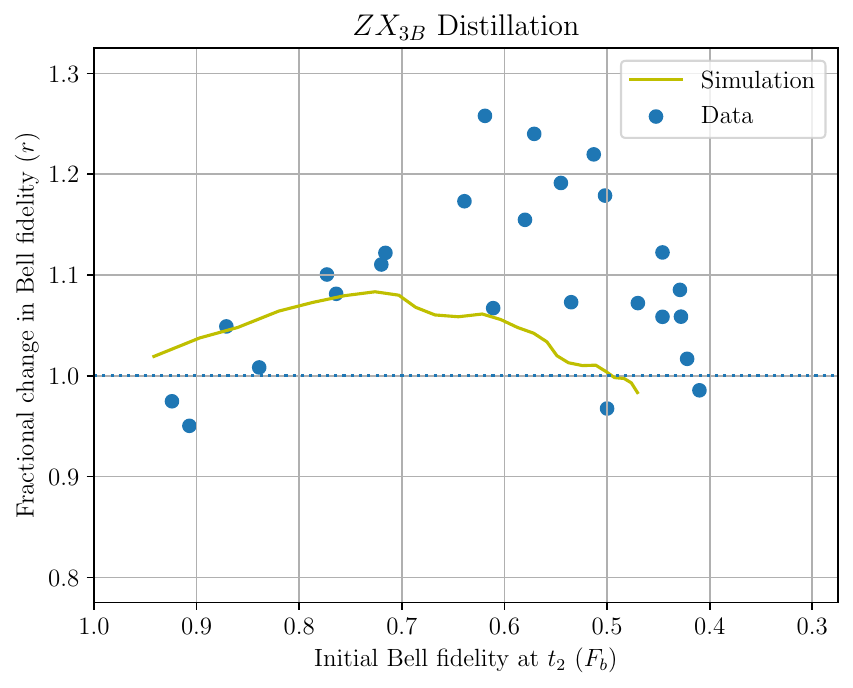}
        \label{fig:3To1CheckRatio}
        }

     \caption{Data and simulation for $ZX_{3B}$ recurrence on
     \texttt{ibm\_kyiv}. \new{In Fig.~\ref{fig:3To1Check2BellAll}
     and~\ref{fig:3To1CheckBellAll} the Bell fidelity for various states is
     plotted against the delay time. States $\rho_1, \rho_2,$ and $\rho_3$
     label the different physically non-local Bell pairs being distilled while
     the state obtained after distillation appears separately.} Solid lines are
     simulations assuming reported noise parameters as described in
     Table.~\ref{table:3To1DevPar_1} and~\ref{table:3To1DevPar_2} of
     App.~\ref{ap:device} \new{while dashed-dotted lines represent experimental
     data.} \new{Figs.~\ref{fig:3To1Check2Ratio} and~\ref{fig:3To1CheckRatio}
     plot the fractional change, $r$ defined in eq.~\eqref{eq:ratio}, against
     the initial Bell fidelity, $F_b$ defined below eq.~\eqref{eq:ratio}.}
     Additional plots in App.~\ref{app:dataDist}, Fig.~\ref{figApp:3To1ExpSim1}
     and Fig.~\ref{figApp:3To1ExpSim2}}
     \label{fig:3To1ExpSim}
\end{figure*}

\subsection{Three Bell distillation experiment with idle noise}
\label{sec:Device3To1}
Using the same technique for experiment and numbers as in the previous sections
we extend the experiment to the three Bell ($ZX_{3B}$) protocol~(circuit
available in App.~\ref{app:circDst}, Fig.~\ref{fig:ExpDist3Bell}) and results
of this are shown in Figs.~\ref{fig:3To1ExpSim}. Again, we get reasonable
agreement with the numerics using just the backend calibration parameters. We
see two cases, in the first~(see Fig.~\ref{fig:3To1Check2BellAll}) we have
small acceptance \new{probability} and no real improvement in the Bell state; here the
noise parameters are quite large. In the second set of data~(see
Fig.~\ref{fig:3To1CheckRatio}) we see some strong improvement from
distillation, but this does not persist well to the highest fidelities and
therefore limits us from distributing higher fidelity entanglement this way. 
In places where we find improvement~(say at e.g. $50 \mu$ s in
Fig.~\ref{fig:3To1CheckBellAll}) one can experience a loss of data, i.e, an
acceptance probability below $50\%$.

\section{Conclusions}

We carried out both a numerical and experimental analysis of some basic
distillation protocols. The simplest numerical analysis in
Sec.~\ref{sec:distill} adds asymmetry among the Bell pairs since the asymmetry
can be inherent in the way multiple Bell pairs are arranged prior to
distillation.  That analysis points at regions in the noise parameter space
where entanglement distillations can provide improvements.  A more involved
analysis using device characterization data, non-Pauli noise channels, and
modelling the echo sequence in our experiment allows us to obtain reasonable
agreement with the experimental data obtained from entanglement distillation.
The noise models indicate how practical difficulties in distillation come from
(1) a natural asymmetry among the Bell pairs being distilled; (2) understanding
the noise which distillation must remove; (3) measurement noise; and (4) fixing
a metric of success for distillation.  When the metric of success is improving
average Bell fidelity over several pairs or removing coherent errors in a
single pair, we find the simplest distillation protocols can provide broad
improvements~(see discussions in Sec.~\ref{sec:glDplExp} and
Sec.~\ref{sec:Device2To1}).  However, for using distilled Bell pairs in modular
computing it is valuable to consider a stricter metric: distill
(physically)~non-local Bell pair with higher fidelity than what can be obtained
using other available means.  This work leaves open the question of broadly
improving upon this metric~\new{(see related theoretical work~\cite{ZangEA25}
in this direction)}.  One way to obtain such improvements using the protocols
discussed here would be to make noise more uniform on the Bell pairs; another
would be to lower the measurement errors on the qubits; while a third would be
to go beyond the types of protocols discussed in this work.

In our analysis, we found it particularly useful to view distillation as
stabilizer checks. This made it `easy' to view different distillation protocols
as tools to catch different types of errors~(which need not be Pauli errors,
and are not necessarily modeled as such here either).  This can help guide the
selection of distillation protocols depending on the source of the dominant
noise in one's error model.

A variety of prior work focusses on the optimal trade-offs between success
probability and Bell fidelity upon distillation over Pauli noise channels. For
practical studies, our work motivates the study of $T_1/T_2$ noise
channels~(these have experimentally available noise parameters) and
benchmarking protocols when starting with unequal bell pairs that incur
measurement errors.

\section{Acknowledgments}
We thank Sam Stein, Michael DeMarco, Andrew Cross and John Smolin for helpful
discussions. This work was supported by the U.S. Department of Energy, Office
of Science, National Quantum Information Science Research Centers, Co-design
Center for Quantum Advantage (C2QA) under contract number DE-SC0012704.

%

\appendix

\section{Notation}
\label{sec:notation}
 
Let $\{\ket{0}, \ket{1}\}$ denote the standard basis of a qubit Hilbert space
$\HC$~(sometimes called the $Z$ basis), $\ket{\pm} := (\ket{0} \pm
\ket{1})/\sqrt{2}$ denote the so-called Hadamard basis, and $H := \dyad{+}{0} +
\dyad{-}{1}$ represent the Hadamard
gate.
Let $X = \dyad{0}{1} + \dyad{0}{1}$, $Y = i\dyad{0}{1} - i \dyad{0}{1}$, and $Z
= \dyad{0}{0} - \dyad{1}{1}$, together with the $2 \times 2 $ identity, $\Ibb$,
denote the Pauli matrices $\PC$.  Any two square matrices $P,Q$ are said to
commute when $PQ = QP$ and anti-commute when $PQ  = - QP$.  Note any two Pauli
matrices either commute or anti-commute.

The qubit X-dephasing~(bit flip) channel,
\begin{equation}
    \DC_q(\rho) = (1-q) \rho + q X \rho X,
    \label{eq:dephChan}
\end{equation}
applies the Pauli $X$ operator with probability $0 \leq q \leq 1/2$.  The qubit
depolarizing channel,
\begin{align}
    \begin{aligned}
    \Lm_p(\rho) &= (1-p) \rho + \frac{p}{3} (X \rho X + Y \rho Y + Z \rho Z), \\
    &= (1-\frac{4}{3}p) \rho + \frac{4p}{3} \Tr(\rho) \frac{\Ibb}{2},
    \label{eq:depolChan}
    \end{aligned}
\end{align}
applies Pauli $X,Y,$ and $Z$ operators with equal probability $p/3$ and the
identity operator $\Ibb$ with probability $1-p$ where $0 \leq p \leq 1$. We use
$\IC$ to denote the identity channel, $\IC(\rho) = \rho$ for all operators
$\rho$.

The fidelity of a two-qubit density operator $\rho$ with the maximally
entangled state,
\begin{equation}
    \ket{\phi} = (\ket{00} + \ket{11})/\sqrt{2},
    \label{eq:BellSt}
\end{equation}
is what we call the Bell fidelity, $F := \mte{\phi}{\rho}$. The Bell
fidelity can be estimated directly from Pauli measurements~(see
App.~\ref{ap:DirBell}). We refer to $1-F$ as the infidelity with the Bell
state~\eqref{eq:BellSt}. The two-qubit CNOT gate with $0^{\text{th}}$ qubit as
control and first qubit as target is $CNOT_{01} = \proj{0} \ot \Ibb + \proj{1}
\ot X$. 

The tensor product of Pauli matrices $\PC$ on $n$-qubits is sometimes called
a Pauli string. We use a compressed notation to represent such a string by
suppressing the identity matrix and using subscript to denote the system label.
For instance with $n=3$ we denote $Z \ot \Ibb \ot X$ by $Z_0 X_2$. 

\section{Direct Bell fidelity estimation}
\label{ap:DirBell}

A two-qubit state, $\rho$, has Bell fidelity,
\begin{equation}
    F = \mte{\phi}{\rho} = \frac{1}{4} (1 + \expc{ZZ}{\rho} + \expc{XX}{\rho} -
    \expc{YY}{\rho})
\end{equation}
where $\expc{N}{\rho} := \Tr(N\rho)$. The Bell fidelity can be computed by
measuring expectation value of $ZZ, XX,$ and $YY$ operators. To compute
$\expc{ZZ}{\rho}$ one may measure each qubit of $\rho$ in the $Z$ basis, obtain
probabilities $p_{ij}$ for measurement outcome corresponding to basis
$\ket{ij}$ and then evaluate $\expc{ZZ}{\rho} = 2(p_{00} + p_{11}) - 1$.

Expectation values for $XX$ and $YY$ can be computed in an analogous manner
by measuring qubits in the $X$ and $Y$ basis instead of $Z$. Thus,
to measure the Bell fidelity this way one does three different measurement
experiments, one for each Pauli measurement basis.
Given a circuit for measuring a qubit in the $Z$ basis one can measure in the
$X$ and $Y$ basis by applying $H$ and $H \cdot S^{\dag}$, respectively, prior to
measuring in the $Z$ basis, where $S = \text{diag}(1,-i)$ is $\sqrt{Z}$.

\section{A general protocol for distillation}
\label{sec:glnDist}

Both recurrence and the $ZX_{3B}$ protocols in Sec.~\ref{sec:distill} of the
main text can be viewed as special cases of those
in~\cite{AmbainisGottesman06,ShorSmolin96}. These distillation protocols can be
viewed in a unified way using a slightly different notation as follows.

Let $\HC_{Ai}$ and $\HC_{Bi}$ each be a qubit Hilbert space, $\HC_i := \HC_{Ai}
\ot \HC_{Bi}$ be a two-qubit Hilbert space, $\HC_{A}:= \otimes_{i=1}^n
\HC_{Ai}$ and $\HC_B:= \otimes_{i=1}^n \HC_{Bi}$ each be $n$-qubit Hilbert
spaces, and $\rho_{AB}$ be a $2n$-qubit state. To distill a single two-qubit
state from $\rho_{AB}$ one applies some unitary $U = U_A \ot U_B$ to
$\rho_{AB}$ then post-selects for agreement on $Z$ measurements made on
$2(n-1)$ qubits. Let the unmeasured system be labeled $0$.  To describe the
post-selected state consider $T = \Ibb_0 \ot L$ where $\Ibb_0$ is the identity
on $\HC_{0}$, $L = \otimes_{i=1}^{n-1}P_i$, and $P_i = \proj{00} + \proj{11}$
is a projector defined on $\HC_i$.  The post-selection succeeds with
probability 
\begin{equation}
    p_a = \Tr(TU\rho U^{\dag}T^{\dag})
    \label{eq:pAccGln}
\end{equation}
and results in a state 
\begin{equation}
    \rho' = \frac{1}{p_a}\Tr_{i \neq 0}(TU\rho U^{\dag}T^{\dag}),
    \label{eq:postSelGln}
\end{equation}
where the partial trace is over all spaces $\HC_i$ except $i=0$. The fidelity
of the post-selected state is
\begin{equation}
     F = \Tr(\rho' \phi).
    \label{eq:postSelF}
\end{equation}

\section{Distilled fidelity calculation}
\label{ap:distCal}

As explained at the end of Sec.~\ref{sec:distill}, the recurrence protocol
post-selects away errors on qubits $2$ and $3$ that anti-commute with $Z_2
Z_3$. This idea can be used to compute the final fidelity $F_a$
in~\eqref{eq:2To1FidDistill} and acceptance probability 
$p_s$ in~\eqref{eq:pAcc2To1}. In Fig.~\ref{fig:distillBasicZ2B}, suppose
\begin{equation}
    \rho_{02} = \IC \ot \PC_{\pB}(\rho) \quad \text{and} \quad
    \rho_{13} = \IC \ot \PC_{\qB}(\rho),
    \label{eq:2To1noiseChannelAp}
\end{equation}
where $\PC_{\pB}(\rho) := p_I \rho +  p_x X \rho X + p_y Y \rho Y + p_z Z \rho
Z$ is a qubit Pauli channel, each $p_i \geq 0$, and $\sum_{i \in \{I,x,y,z\}}
p_i = 1$. 
The collection of the Pauli errors not caught by recurrence are detailed in
Table~\ref{table:1}. In this table, the first column denotes errors on qubits
$2$ and $3$ that commute with $Z_2 Z_3$, and are thus not post-selected away by
recurrence, the second column denotes the probability of these errors, and the
third column denotes the effective error after post-selection on qubit labelled
$2$.  Setting $p_x = p_y = p_z = p/3$ and $q_x = q_y = q_z = q/3$ and summing
the entries in the second column gives the acceptance probability
$p_s$~\eqref{eq:pAcc2To1} while summing those entries in the second column
where the effective error is $I_2$ and normalizing by $p_s$ gives the Bell
fidelity of the distilled state $F_a$~\eqref{eq:2To1FidDistill}. 

\begin{table}[h!]
    \centering
\begin{tabular}{||c c c||} 
 \hline
 Error & Probability & Transformed Error \\ [0.5ex] 
 \hline\hline
 $\Ibb$ & $p_I q_I$ & $\Ibb_2$ \\ 
 \hline
 $Z_3$ & $p_I q_z$ & $Z_2$ \\
 \hline
 $X_2X_3$ & $p_xq_x$ & $X_2$ \\
 \hline
 $X_2Y_3$ & $p_xq_y$ & $Y_2$ \\
 \hline
 $Y_2X_3$ & $p_yq_x$ & $Y_2$ \\ 
 \hline
 $Y_2Y_3$ & $p_yq_y$ & $X_2$ \\
 \hline
 $Z_2$ & $p_zq_I$ & $Z_2$ \\ 
 \hline
 $Z_2Z_3$ & $p_zq_z$ & $\Ibb_2$ \\ [1ex] 
 \hline
\end{tabular}
    \caption{Accepted errors by recurrence and their effect: Pauli error not
    post-selected away by recurrence~(see Sec.~\ref{sec:rec}), their
    probability, and their effect on the final distilled Bell pair.}
\label{table:1}
\end{table}

For the $ZX_{3B}$ distillation protocol, we may derive the acceptance
probability $p_s$ in~\eqref{eq:3To1ps} and Bell fidelity upon acceptance $F_b$
in~\eqref{eq:3To1FidDistill} using a procedure analogous to the one above.
Suppose in Fig.~\ref{fig:3To1DepolUneq} 
\begin{align}
    \begin{aligned}
        \rho_{03} &= \IC \ot \PC_{\pB}(\rho), \\
        \rho_{14} &= \IC \ot \PC_{\qB}(\rho), \quad \text{and}\\ 
        \rho_{25} &= \IC \ot \PC_{\rB}(\rho).
    \end{aligned}
    \label{eq:3To1noiseChanTwo}
\end{align}
then we list the collection of errors not caught by the $ZX_{3B}$ distillation
protocol in Table~\ref{table:2}.  In this table, the first, second and third
column represent the accepted errors~(those which commute with both $Z_3 Z_4
Z_5$ and $X_4 X_5$), their probability and their effect on qubit $3$ after
qubits $4$ and $5$ are measured, respectively.
Setting $p_x = p_y = p_z = r_x = r_y = r_z = p/3$, and $q_x = q_y = q_z = q/3$
and summing the entries in the second column gives the acceptance probability,
$p_s$ in~\eqref{eq:3To1ps}, while summing the entries in the second column
corresponding to no error~($\Ibb$) in the third and normalizing by $p_s$ gives
the fidelity after distillation, $F_a$ in~\eqref{eq:3To1FidDistill}.

\begin{table}[h!]
    \centering
\begin{tabular}{||c c c||} 
 \hline
 Error & Probability & Transformed Error \\ [0.5ex] 
 \hline\hline
 $\Ibb$ & $p_I q_I r_I$ & $\Ibb_3$ \\ 
 \hline
 $X_4 X_5$ & $p_I q_x r_x$ & $\Ibb_3$ \\
 \hline
 $Z_4 Z_5$ & $p_I q_z r_z$ & $Z_3$ \\
 \hline
 $Y_4 Y_5$ & $p_I q_y r_y$ & $Z_3$ \\
 \hline
 $X_3 X_5$ & $p_x q_I r_x$ & $X_3$ \\ 
 \hline
 $X_3 X_4$ & $p_x q_x r_I $ & $X_3$ \\
 \hline
 $X_3 Z_4 Y_5$ & $p_x q_z r_y$ & $Y_3$ \\ 
 \hline
 $X_3 Y_4 Z_5$ & $p_x q_y r_z$ & $Y_3$ \\ 
 \hline
 $Z_3$ & $p_z q_I r_I$ & $Z_3$ \\ 
 \hline
 $Z_3 X_4 X_5$ & $p_z q_x r_x$ & $Z_3$ \\ 
 \hline
 $Z_3 Z_4 Z_5$ & $p_z q_z r_z$ & $\Ibb_3$ \\ 
 \hline
 $Z_3 Y_4 Y_5$ & $p_z q_y r_y$ & $\Ibb_3$ \\ 
 \hline
 $Y_3 X_5$ & $p_y q_I r_x$ & $Y_3$ \\ 
 \hline
 $Y_3 X_4$ & $p_y q_x r_I$ & $Y_3$ \\ 
 \hline
 $Y_3 Z_4 Y_5$ & $p_y q_z r_y$ & $X_3$ \\ 
 \hline
 $Y_3 Y_4 Z_5$ & $p_y q_y r_z$ & $X_3$ \\ [1ex] 
 \hline
\end{tabular}
    \caption{Accepted errors by $ZX_{3B}$ and their effect: Pauli error not
    post-selected away by the $ZX_{3B}$ protocol~(see Sec.~\ref{sec:ZZZIXX}),
    their probability, and their effect on the final distilled Bell pair.}
\label{table:2}
\end{table}

\section{Device Overview}
\label{ap:device}

Experiments were carried out on IBM's fixed-frequency transmon
superconducting processor, \texttt{ibm\_kyiv}. This is a 127 qubit
chip with qubits arranged in a heavy-hexagonal lattice which reduces
cross-talk with reasonable overhead in circuit layout mapping. This
processor is from IBM's \texttt{Eagle} processor family which use the
echoed cross-resonance gate for its entangling gate and features
multiplexed readout. To support the higher qubit count, the chip
features multi-layer wiring with care taken to reduce the effects of
quantum and classical cross-talk. This architecture results in a median $T_1$
and $T_2$ of 276 \unit{\us} and 122 \unit{\us} respectively and median ECR gate
error and readout error of \num{1.1e-2} and \num{6.0e-3} respectively.
Tables~\ref{table:2To1DevPar_Z}, ~\ref{table:2To1DevPar_X},
~\ref{table:3To1DevPar_1}, and~\ref{table:3To1DevPar_2} give error parameters
for specific qubits on which distillation experiments and simulation were
discussed in the main text.

\begin{table}[h!]
    \centering
\begin{tabular}{||c c c c ||} 
 \hline
    Qubit & $T_1~(\mu s)$ & $T_2~(\mu s)$ & Measurement Error \\ [0.5ex] 
 \hline\hline
    0 & 257.944 & 323.573 & 6.5 $\times 10^{-3}$\\ 
    \hline
    1 & 477.815 & 224.595 & 9.1 $\times 10^{-3}$\\
    \hline
    2 & 263.123 & 123.047 & 4.3 $\times 10^{-3}$\\
    \hline
    3 & 260.839 & 232.639 & 4.6 $\times 10^{-3}$\\ 
    \hline
\end{tabular}

    \vspace{1em}

\begin{tabular}{||c c c c ||} 
 \hline
    Qubit 1 & Qubit 2 & $ZZ$ Rate~(Hz) & ECR Error \\ [0.5ex] 
    \hline\hline
     0 & 1 & -52860.4 & 7.75153 $\times 10^{-3}$\\ 
     \hline
     1 & 2 & -55319.3 & 10.3203  $\times 10^{-3}$\\ 
     \hline
     2 & 3 & -45908   & 4.2953  $\times 10^{-3}$\\ 
   \hline
\end{tabular}
    \caption{Single qubit and two-qubit error rates from \texttt{ibm\_kyiv}
    last updated on 2024-05-14 at 07:45:06 UTC used for simulation of
    recurrence $Z_{2B}$.  Measurement delay is $1.24 \mu s$.  The CNOT error
    rate is taken to be the ECR error rate and kept the same when role of
    target and control qubits is reversed.}
\label{table:2To1DevPar_Z}
\end{table}

\begin{table}[h!]
    \centering
\begin{tabular}{||c c c c ||} 
 \hline
    Qubit & $T_1~(\mu s)$ & $T_2~(\mu s)$ & Measurement Error \\ [0.5ex] 
 \hline\hline
    0 & 276.892 & 312.245  & 2.5 $\times 10^{-3}$\\ 
    \hline
    1 & 512.747 & 218.116  & 2.6 $\times 10^{-3}$\\
    \hline
    2 & 236.636 & 98.102 & 4.2 $\times 10^{-3}$\\
    \hline
    3 & 330.719 & 232.639 & 11.2 $\times 10^{-3}$\\ 
    \hline
\end{tabular}

    \vspace{1em}

\begin{tabular}{||c c c c ||} 
 \hline
    Qubit 1 & Qubit 2 & $ZZ$ Rate~(Hz) & ECR Error \\ [0.5ex] 
    \hline\hline
     0 & 1 & -52860.4 &  4.43472  $\times 10^{-3}$\\ 
     \hline
     1 & 2 & -55319.3 &  8.10392  $\times 10^{-3}$\\ 
     \hline
     2 & 3 & -45908   & 4.03714  $\times 10^{-3}$\\ 
   \hline
\end{tabular}
    \caption{Single qubit and two-qubit error rates from \texttt{ibm\_kyiv} last updated
    on 2024-05-15 at 16:05:02 UTC used for simulation of recurrence $X_{2B}$.
    Measurement delay is $1.24 \mu s$.  The CNOT error rate is taken to be the
    ECR error rate and kept the same when role of target and control qubits is
    reversed.}
\label{table:2To1DevPar_X}
\end{table}

\begin{table}[h!]
    \centering
\begin{tabular}{||c c c c ||} 
 \hline
    Qubit & $T_1~(\mu s)$ & $T_2~(\mu s)$ & Measurement Error \\ [0.5ex] 
 \hline\hline
  3 & 410.738 & 232.639 & 4.2 $\times 10^{-3}$ \\ 
  \hline
  4 & 429.253 & 152.675 & 16.9 $\times 10^{-3}$ \\
  \hline
  5 & 365.064 & 384.404 & 6.8 $\times 10^{-3}$ \\
  \hline
  6 & 294.473 & 146.729 & 1.6 $\times 10^{-3}$ \\
  \hline
  7 & 339.754 & 371.239 & 3.4 $\times 10^{-3}$ \\
  \hline
  8 & 458.343 & 259.075 & 2.7 $\times 10^{-3}$ \\ 
  \hline
  59 & 269.232 & 78.7512 & 2.4  $\times 10^{-3}$ \\
  \hline
  60 & 273.893 & 285.543 & 6   $\times 10^{-3}$ \\
  \hline
  61 & 318.205 & 152.633 & 7.9  $\times 10^{-3}$ \\
  \hline
  62 & 255.428 & 25.5405 & 23.6  $\times 10^{-3}$ \\
  \hline
  63 & 275.333 & 115.334 & 7.3  $\times 10^{-3}$ \\
  \hline
  64 & 231.769 & 47.6543 & 3.8  $\times 10^{-3}$ \\  [1ex]
 \hline
\end{tabular}
    \caption{Single qubit error rates from \texttt{ibm\_kyiv} last updated on 2024-05-17
    at 05:54:35 UTC used for simulation of the three bell experiment.
    Measurement delay is $1.24 \mu s$}
\label{table:3To1DevPar_1}
\end{table}

\begin{table}[h!]
    \centering
\begin{tabular}{||c c c c ||} 
 \hline
    Qubit 1 & Qubit 2 & $ZZ$ Rate~(Hz) & CNOT Error \\ [0.5ex] 
 \hline\hline
      3 &         4 &  -48982   & 8.49846 $\times 10^{-3}$ \\ 
      \hline
      4 &         5 &  -39813.1 & 11.4476  $\times 10^{-3}$ \\ 
      \hline
      5 &         6 &  -76347.7 & 8.78052 $\times 10^{-3}$ \\ 
      \hline
      6 &         7 &  -57303.5 & 16.352   $\times 10^{-3}$ \\ 
      \hline
      7 &         8 &  -40264.1 & 8.80976 $\times 10^{-3}$ \\ 
      \hline
     59 &        60 &  -127831  & 13.0765  $\times 10^{-3}$ \\ 
      \hline
     60 &        61 &  -38618.7 & 7.85857 $\times 10^{-3}$ \\ 
      \hline
     61 &        62 &  -57210.3 & 13.1797  $\times 10^{-3}$ \\ 
      \hline
     62 &        63 &  -55771.1 & 13.9869  $\times 10^{-3}$ \\ 
      \hline
     63 &        64 &  -40636.1 & 6.07738 $\times 10^{-3}$ \\ [1ex]
      \hline
\end{tabular}
    \caption{Two qubit error rate and $ZZ$ rate from \texttt{ibm\_kyiv} last updated on
    2024-05-17 at 05:54:35 UTC used for simulation of the three bell
    distillation protocol. The CNOT error rate is taken to be the same when
    role of target and control qubits is reversed.} 
    \label{table:3To1DevPar_2}
\end{table}

\section{$T_1$/$T_2$ Channel}
\label{ap:T1T2chan}

Noise on a superconducting qubit can be described using a damping-dephasing
channel~\cite{AliferisBritoEA09} that may be expressed
as~\cite{SiddhuAbdelhadiEA24},
\begin{equation}
    \MC(\rho) = \sum_i O_i \rho O_i^{\dag},
    \label{eq:dampDeph}
\end{equation}
where $O_0 = \sqrt{1-p}(\proj{0}+\sqrt{1-g}\proj{1})$, $O_1= \sqrt{g}
\ket{0}\bra{1}$, $O_2 = \sqrt{p}(\proj{0}-\sqrt{1-g}\proj{1})$, $0 \leq p \leq
1/2$ represents dephasing and $0 \leq g \leq 1$ represents damping probability.
The channel maps an input density operator with Bloch vector $\rB = (x,y,z)$ to
an output with Bloch coordinates $\big( (1-2p)\sqrt{1-g}x,  (1-2p)\sqrt{1-g}y,
(1-g)z + g \big)$.  When the output coordinates are parametrized as
$(e^{-t/T_2}x, e^{-t/T_2}y, e^{-t/T_1}z + 1 - e^{-t/T_1} )$ then 
\begin{equation}
    g = 1 - e^{-t/T_1}, \quad 
    p = (1 - e^{-t(1/T_2 - 1/2T_1)})/2,
    \label{eq:gpT1T2}
\end{equation}
and $2T_1 \geq T_2$
When $p = 0$ or $T_2 = 2T_1$, $\MC$ is an amplitude damping channel mapping
$\ket{1}$ to $\ket{0}$ with probability $g$. When $g = 0$, or $T_1 = \infty$,
$\MC$ is a pure dephasing channel that applies a $Z$ error with probability
$p$.
If a qubit with fixed $T_1$ and $T_2$ parameters waits idle for a time $t$ then
channel modelling noise on this qubit, $\MC_t$, is the $\MC$ channel with
paramters $g$ and $p$ depend on $t, T_1, $ and $T_2$, as indicated
in~\eqref{eq:gpT1T2}.

\newpage
\onecolumngrid

\section{Circuits for noise and distillation}
\label{app:circDst}
Here we present the circuits used in the distillation protocols of the main
text, as well as the modifications to those circuits that enabled us to study
them under various types of noise.
\begin{figure*}[htp]
     \subfloat[][Two Bell distillation circuit with global depolarizing noise.
     ]{
         \includegraphics[clip,width=.45\columnwidth]{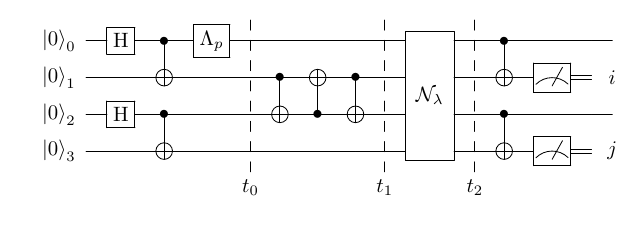} 
     \label{fig:TwoToOneGl}}
     \qquad
     \subfloat[][Three Bell distillation circuit with global depolarizing noise.]{
         \includegraphics[clip,width=.45\columnwidth]{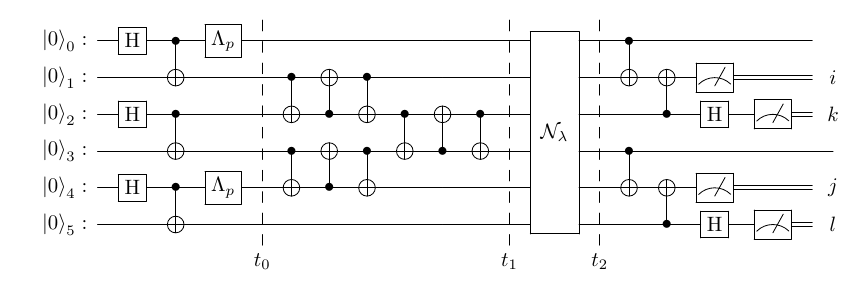} 
     \label{fig:ThreeToOneGl}}
     \caption{Circuits for simulation of entanglement distillation with global
     depolarizing channel $\NC_\lm$ replacing the local depolarizing noise between
     $t_1$ and $t_2$ barriers in Fig.~\ref{fig:circsDist} of the main text.}
     \label{fig:circsGl}
\end{figure*}

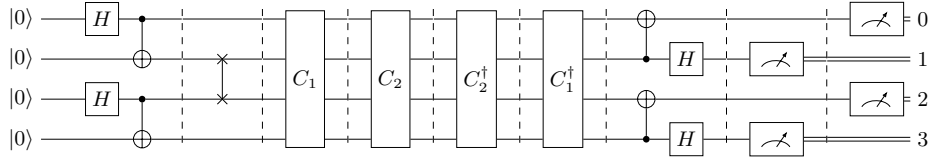
\begin{figure}[htp]
\centering
  \resizebox{.7\columnwidth}{!}{
\begin{tikzpicture}
\begin{yquant}[register/minimum height=6pt, register/minimum depth=6pt]
qubit  {$\ket{0}_0$} a;
qubit  {$\ket{0}_1$} b;
qubit  {$\ket{0}_2$} c;
qubit  {$\ket{0}_3$} d;
hspace {16pt} -;
h a;
h c;
cnot b | a;
cnot d | c;
barrier (a, b , c, d);
swap (b, c);
barrier (a, b , c, d);
box {$C_1$} (a, b , c, d);
barrier (a, b , c, d);
box {$C_2$} (a, b , c, d);
barrier (a, b , c, d);
box {$C_2^\dagger$} (a, b , c, d);
barrier (a, b , c, d);
box {$C_1^\dagger$} (a, b , c, d);
barrier (a, b , c, d);
cnot b | a;
cnot d | c;
barrier (a, b , c, d);
measure b;
measure d;
output {$i$} b;
output {$j$} d;
barrier (a, b , c, d);
measure a;
measure c;
output {} a;
output {} c;
\end{yquant}
\end{tikzpicture}
}
    \caption{Circuit for implementing recurrence under global depolarizing
    noise using mirror Clifford layers. \new{Bell pairs are created before the
    first barrier~(dotted vertical line), then qubits $1$ and $2$ are swapped
    to create physically non-local Bell pairs, next two layers of Clifford 
    circuits $C_1, C_2$ and their inverse are applied, and followed by a
    recurrence protocol for distillation and measurement of all qubits.}}
    \label{fig:Layer}
\end{figure}

\begin{figure*}[htp]
    \centering
    \subfloat[][Two bell creation with idling noise.]
    {\includegraphics[width=.8\columnwidth]{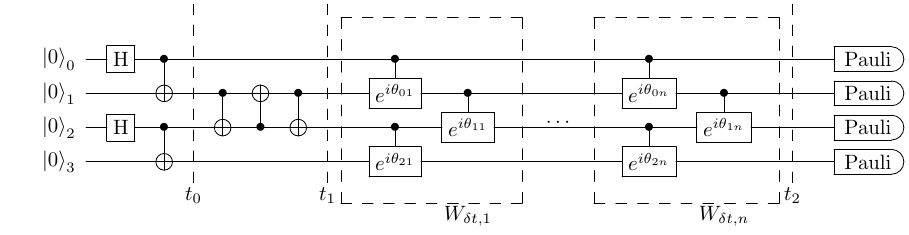} 
    \label{fig:DeviceExp}}
    \qquad
    \subfloat[][$Z_{2B}$ recurrence protocol with idling noise]
    {\includegraphics[width=.8\columnwidth]{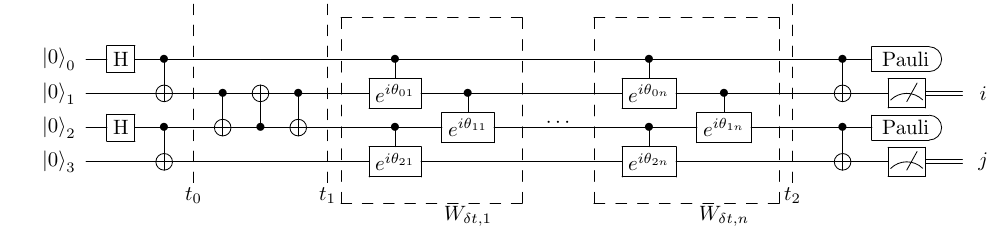}
    \label{fig:DeviceExpDist}}
    \caption{Circuit(s) for the $Z_{2B}$ recurrence protocol with
    idling noise. In both circuits, two Bell pairs are created prior to time
    label $t_0$, between $t_0$ and $t_1$ the first and second qubits are
    swapped and a delay is added from $t_1$ to $t_2$.
    The delay inteval $t_2  - t_1$ is parametrized as $n \delta t$ where
    $\delta t$ is variable and fixed integer $n$ is the number of times the
    $ZZ$ sequence $W_{\delta t, k}$ is applied.  Control phase angles in each
    sequence are $\th_{jk} = f_k 2 \pi \omega_j \delta t/n$ where $ 1 \leq k
    \leq n$, $0 \leq j \leq 2$, $f_k = 1$ if $k<n/2$ and $-1$ otherwise,
    $\omega_j$ is the $ZZ$ frequence between qubits $j$ and $j+1$.
    In Fig.~\ref{fig:DeviceExp} after time $t_2$ Pauli basis measurements are
    carried out on each qubit to determine the Bell fidelity~(see
    App.~\ref{ap:DirBell} for details).
    In Fig.~\ref{fig:DeviceExpDist} after $t_2$, $ZZ$ measurements are carried
    out on qubits $1$ and $3$ while qubits $0$ and $2$ are measured in the
    Pauli basis for Bell fidelity estimation.}
    \label{fig:DeviceExpDist2Bell}
\end{figure*}

\begin{figure*}[htp]
    \centering
    \subfloat[][Three bell creation with idling noise.]
    {\includegraphics[width=.8\columnwidth]{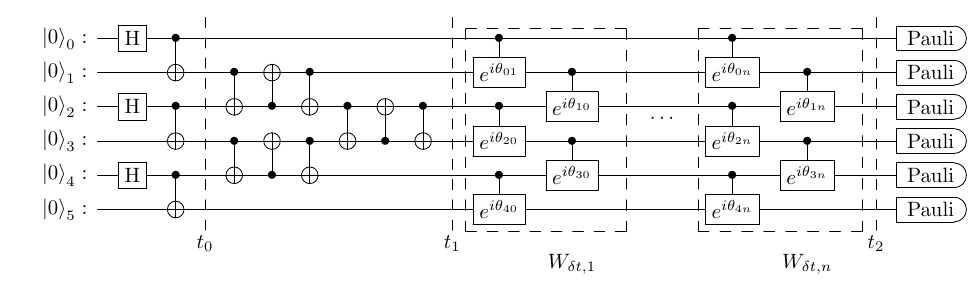} 
    \label{fig:DeviceExp3Bell}}
    \qquad
    \subfloat[][$ZX_{3B}$ distillation with idling noise]
    {\includegraphics[width=.8\columnwidth]{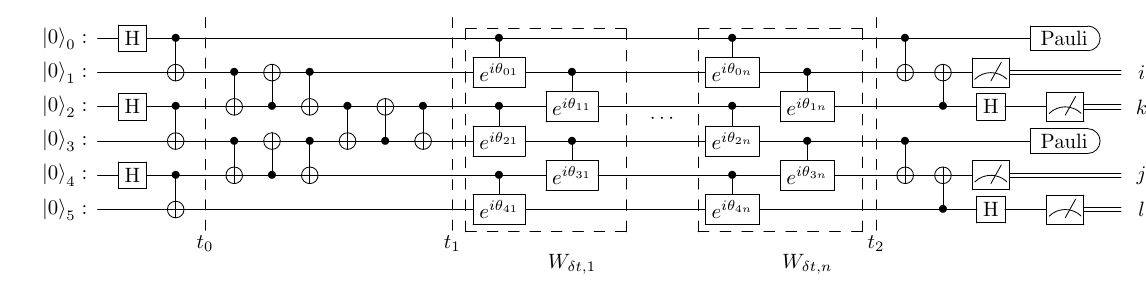}
    \label{fig:DeviceExpDist3Bell}}
    \caption{Circuit(s) for $ZX_{3B}$ distillation with idling noise. In both
    circuits, three Bell pairs are created prior to time label $t_0$, between
    $t_0$ and $t_1$ the Bell pairs are swapped~(see Sec.~\ref{sec:Local3To1}
    for details) and a delay is added from $t_1$ to $t_2$~(in a mannar
    analogous to that in Fig.~\ref{fig:DeviceExpDist2Bell}).
    In Fig.~\ref{fig:DeviceExp3Bell} after time $t_2$ Pauli basis measurements
    are carried out on each qubit to determine the Bell fidelity~(see
    App.~\ref{ap:DirBell} for details).
    In Fig.~\ref{fig:DeviceExpDist3Bell} after $t_2$, $ZZZ$ amd $IXX$
    measurements are carried out~(as explained in Sec.~\ref{sec:ZZZIXX}) while
    qubits $0$ and $3$ are measured in the Pauli basis for Bell fidelity
    estimation.}
    \label{fig:ExpDist3Bell}
\end{figure*}

\newpage

\section{Additional plots from simulation and experiments}
\label{app:dataDist}
Here we present plots generated using both simulations and experiments of
entanglement distillation protocols discussed in the main text. Plots from
simulation extend the ones in the main text, and those with experimental data
separate out the Bell fidelity already presented and also show the acceptance
probability.
\begin{figure*}[htp]
    \centering
     \subfloat[][\new{Bell pairs are prepared
     with equal Bell Fidelity.}]{
         \includegraphics[clip,width=.45\columnwidth]{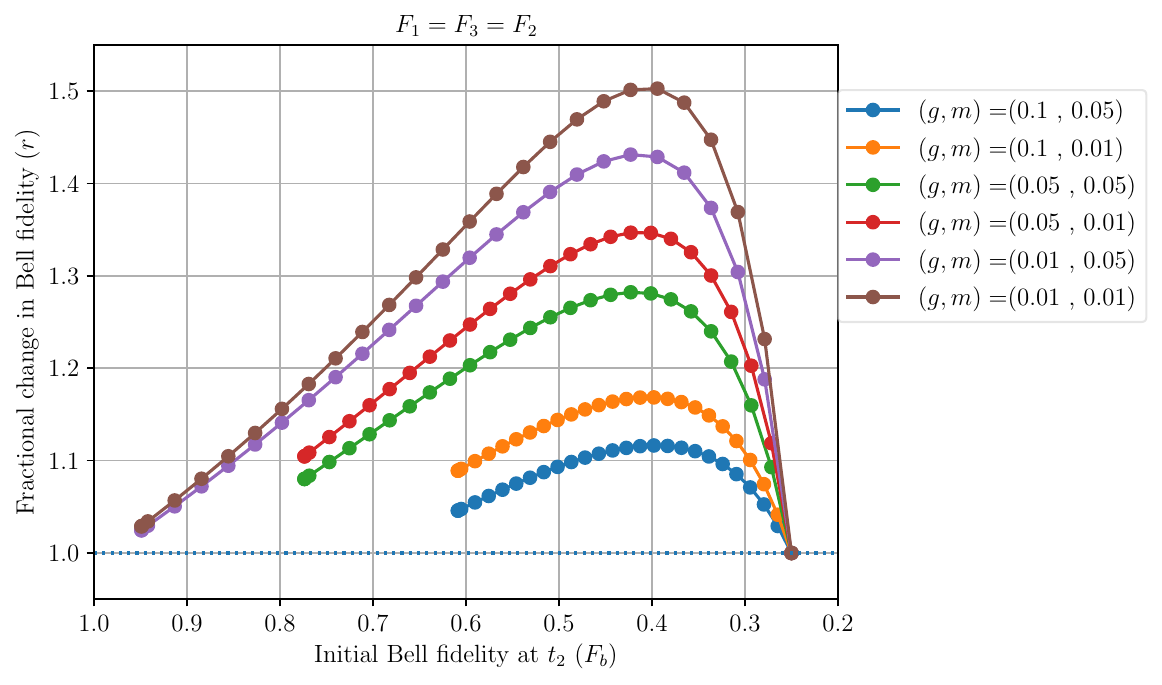}
     \label{fig:3To1PlotGl3}
     }
     \qquad
     \subfloat[][\new{Bell pairs are prepared
     with unequal Bell fidelity, the first and third have the same Bell fidelity which
     is $9 \%$ less than that of the second Bell pair.}]{
         \includegraphics[clip,width=.45\columnwidth]{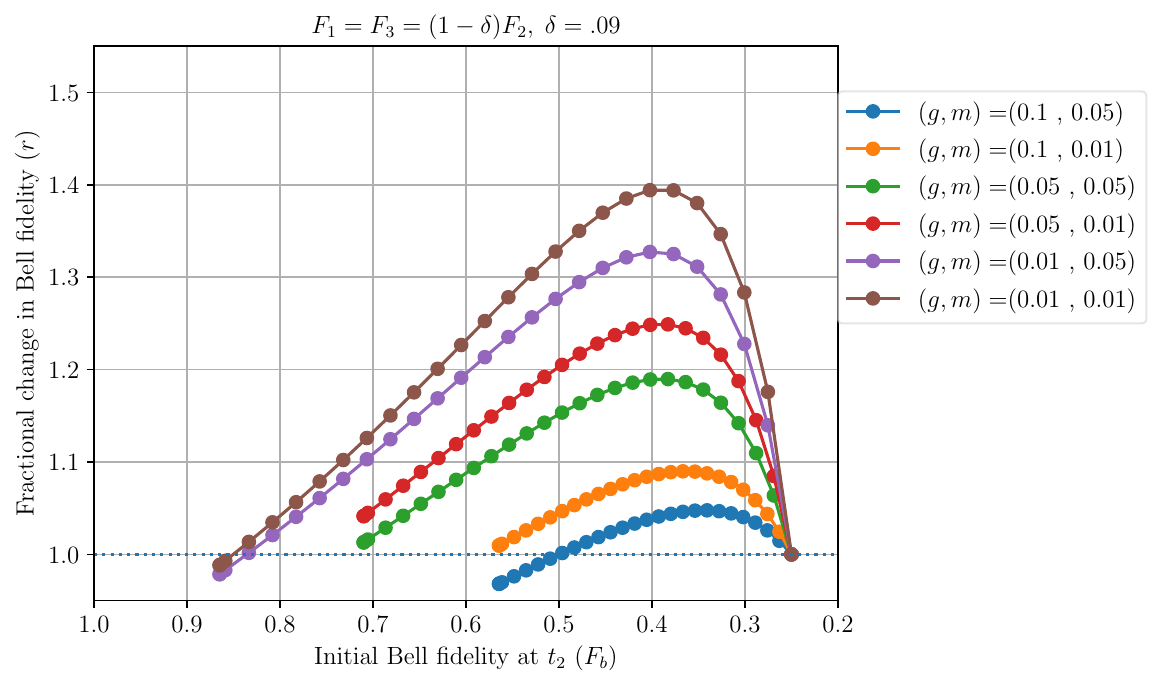}
     \label{fig:3To1PlotGl4}
     }
     \caption{
     \new{Results from simulation of recurrence $ZX_{3B}$ with global
     depolarizing noise~(for circuit see Fig.~\ref{fig:ThreeToOneGl} in
     App.~\ref{app:circDst}). Plots~\eqref{fig:3To1PlotGl3}
     and~\eqref{fig:3To1PlotGl4} show fractional change in Bell fidelity,
     $r$ defined in eq.~\eqref{eq:ratio}, plotted against initial Bell
     fidelity, $F_b$~(defined below eq.~\eqref{eq:ratio}), for various gate and
     measurement errors. The region of the plot above $r=1$ indicates where the
     noisy distillation circuit is beneficial.}}
     \label{fig:3To1PlotGlobal}
\end{figure*}
\begin{figure*}[htp]
    \begin{minipage}{.6\linewidth} 
    \centering
     \subfloat[][First Bell pair]{
         \includegraphics[clip,width=.5\columnwidth]{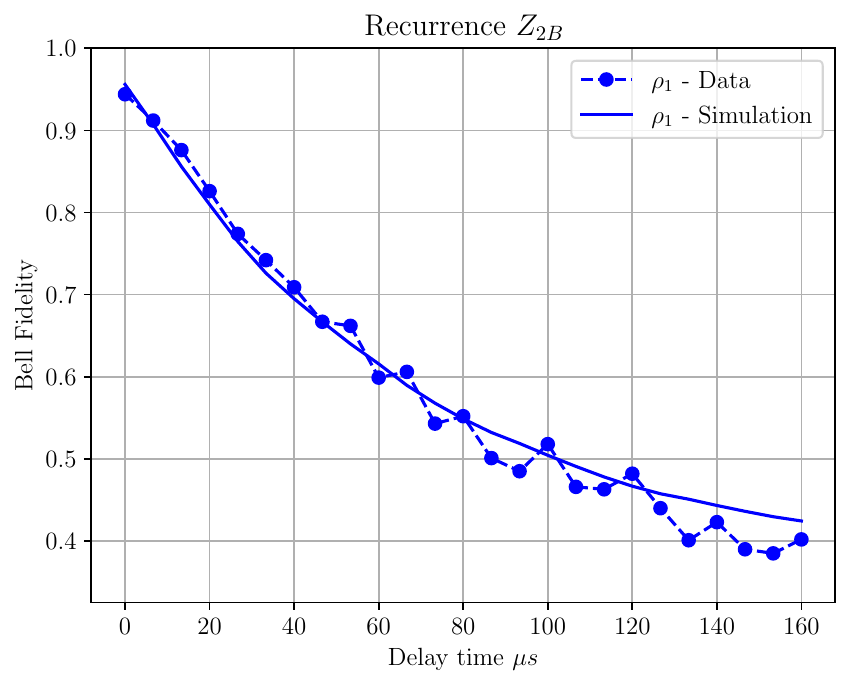}
     \label{fig:XCheckBell1}
     }
     \subfloat[][Second Bell pair]{
         \includegraphics[clip,width=.5\columnwidth]{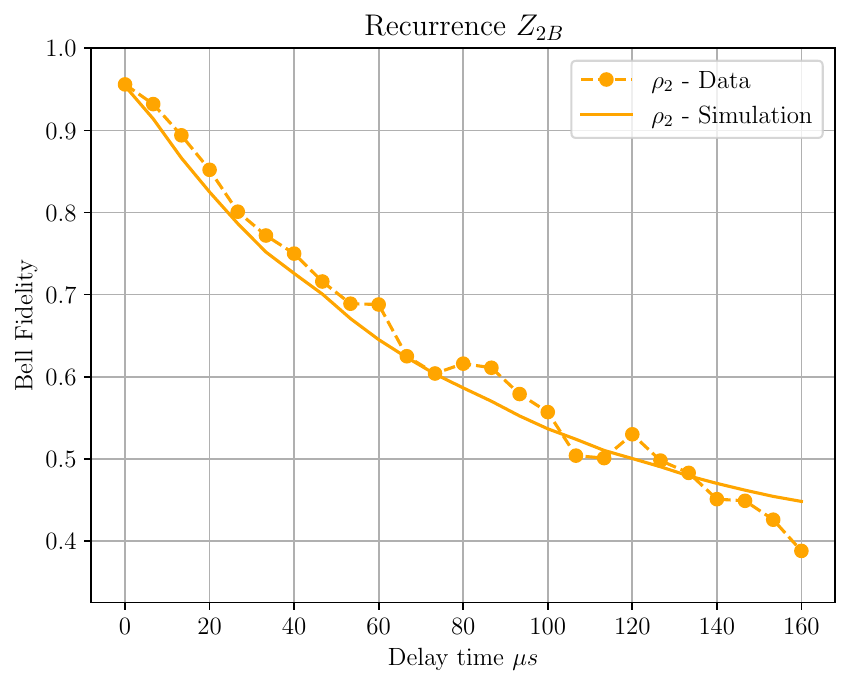}
     \label{fig:XCheckBell2}
     }
    \end{minipage}
    \centering
    \begin{minipage}{.6\linewidth}
     \subfloat[][Distilled Bell pair]{
         \includegraphics[clip,width=.5\columnwidth]{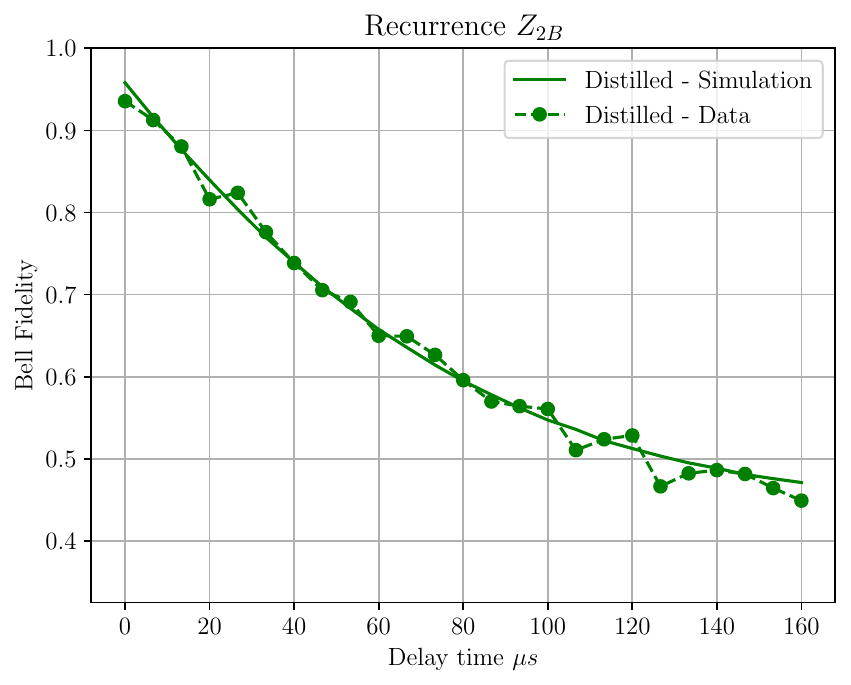}
     \label{fig:XCheckBellD}
     }
     \subfloat[][Acceptance probability]{
         \includegraphics[clip,width=.5\columnwidth]{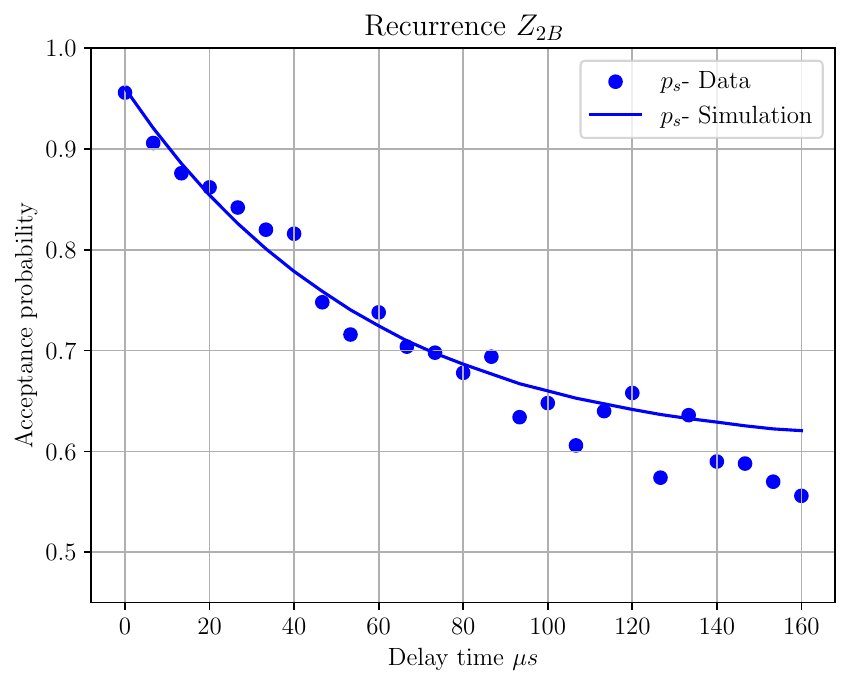}
     \label{fig:XCheckPacc}
     }
    \end{minipage}
     \caption{$Z_{2B}$ Recurrence distillation on qubits [0,1,2,3] of
     \texttt{ibm\_kyiv} (points). Here simulations fit the data well.  Solid
     lines are simulations assuming reported noise parameters as described in
     Table.~\ref{table:2To1DevPar_Z} of App.~\ref{ap:device}.
     \new{Figs.~\ref{fig:XCheckBell1}, ~\ref{fig:XCheckBell2},
     and~\ref{fig:XCheckBellD} plot Bell fidelity against delay time for the
     respective physically non-local state labelled in the figure and
     Fig.~\ref{fig:XCheckPacc} shows the acceptance probability versus delay
     time.}}
     \label{figApp:2To1ExpSim}
\end{figure*}
\begin{figure*}[htp]
    \begin{minipage}{.6\linewidth} 
    \centering
     \subfloat[][First Bell pair]{
         \includegraphics[clip,width=.5\columnwidth]{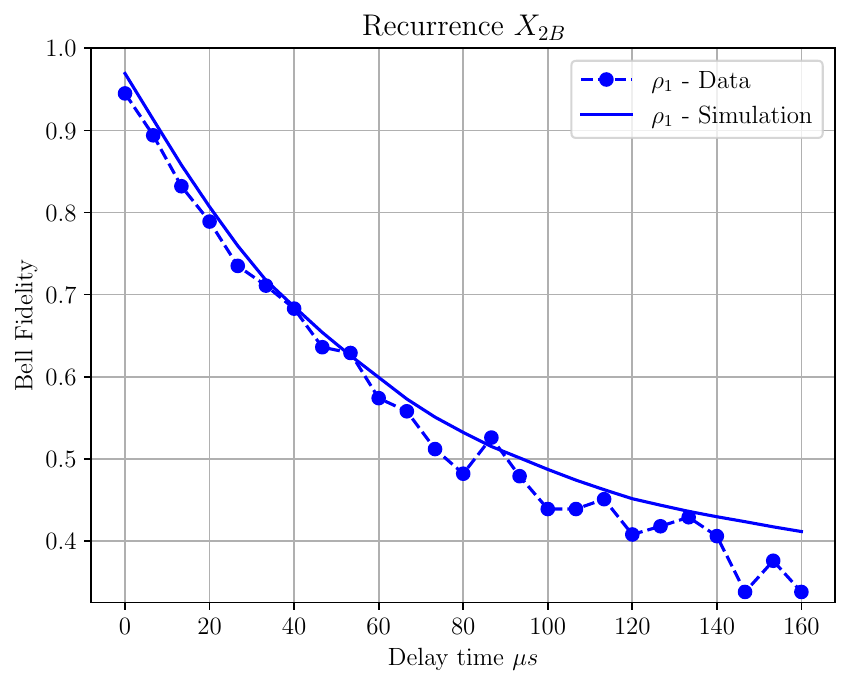}
     \label{fig:ZCheckBell1}
     }
     \subfloat[][Second Bell pair]{
         \includegraphics[clip,width=.5\columnwidth]{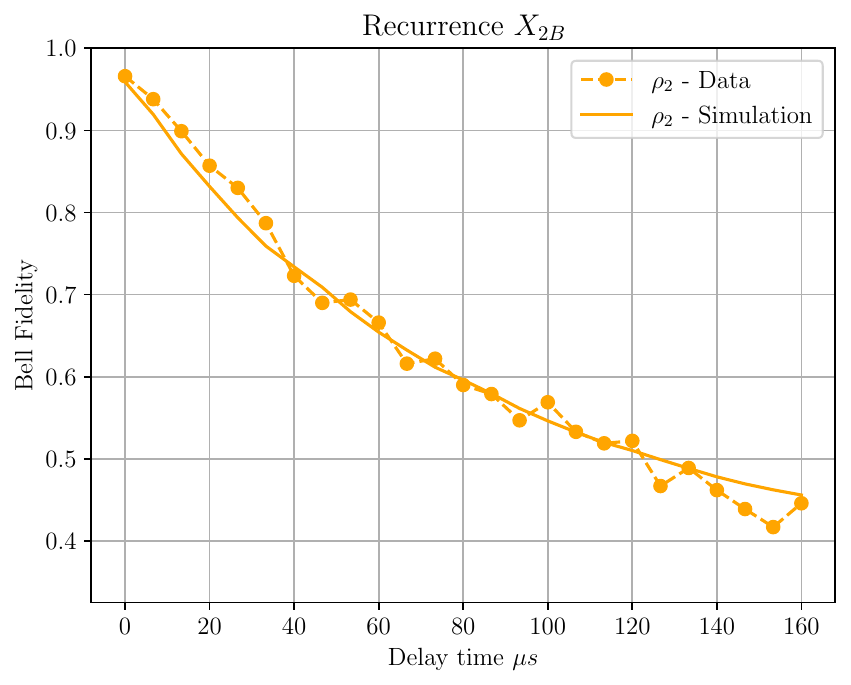}
     \label{fig:ZCheckBell2}
     }
    \end{minipage}
    \centering
    \begin{minipage}{.6\linewidth}
     \subfloat[][Distilled Bell pair]{
         \includegraphics[clip,width=.5\columnwidth]{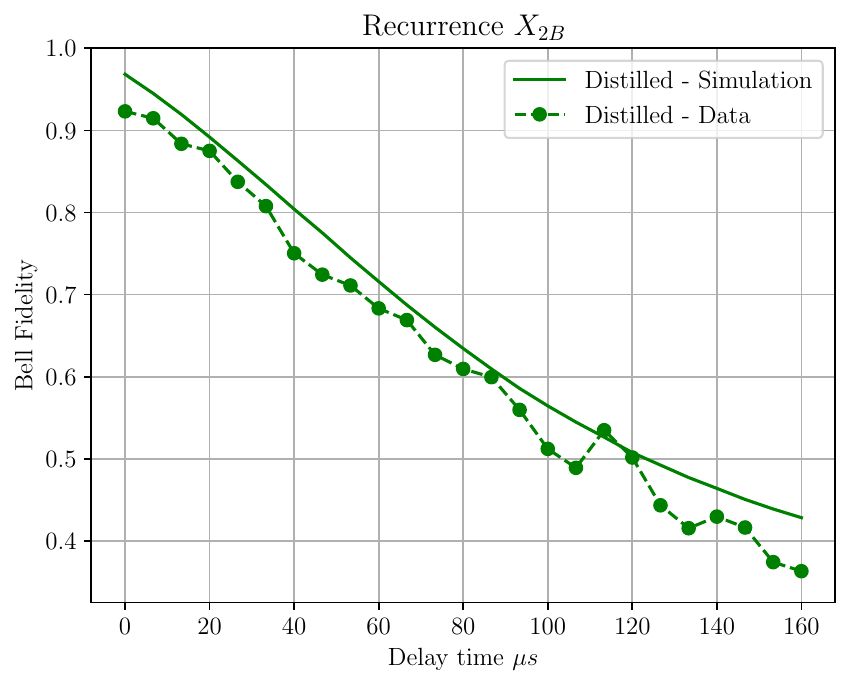}
     \label{fig:ZCheckBellD}
     }
     \subfloat[][Acceptance probability]{
         \includegraphics[clip,width=.5\columnwidth]{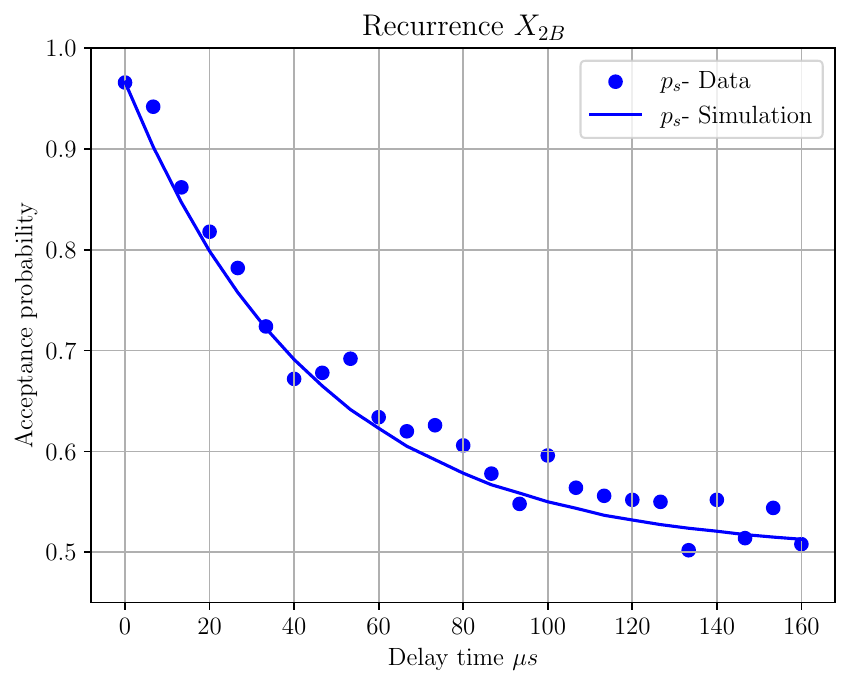}
     \label{fig:ZCheckPacc}
     }
    \end{minipage}
     \caption{$X_{2B}$ \new{Recurrence} distillation on qubits [0,1,2,3] of
     \texttt{ibm\_kyiv} (points). \new{Here simulations fit the data well.}
     Solid lines are simulations assuming reported noise parameters as
     described in Table.~\ref{table:2To1DevPar_X} of App.~\ref{ap:device}.
     \new{Figs.~\ref{fig:ZCheckBell1}, ~\ref{fig:ZCheckBell2},
     and~\ref{fig:ZCheckBellD} plot Bell fidelity against delay time for the
     respective physically non-local state labelled in the figure and
     Fig.~\ref{fig:ZCheckPacc} shows the acceptance probability versus delay
     time.}}
     \label{figApp:2To1ExpSimZCheck}
\end{figure*}

\begin{figure*}[htp]
    \begin{minipage}{.9\linewidth} 
    \centering
     \subfloat[][First Bell pair]{
         \includegraphics[clip,width=.4\columnwidth]{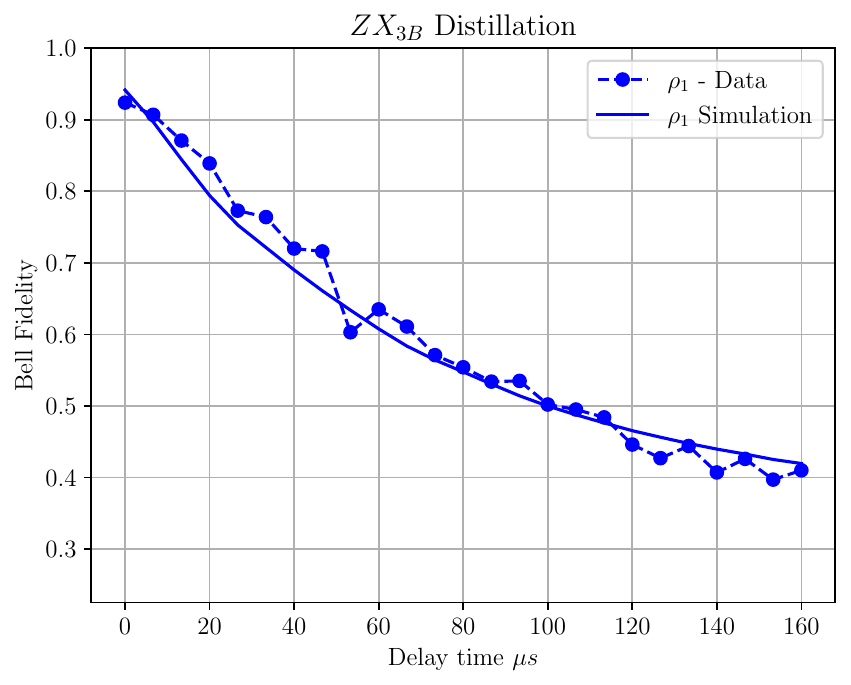}
     \label{fig:3To1Check2Bell1}
     }
     \subfloat[][Second Bell pair]{
         \includegraphics[clip,width=.4\columnwidth]{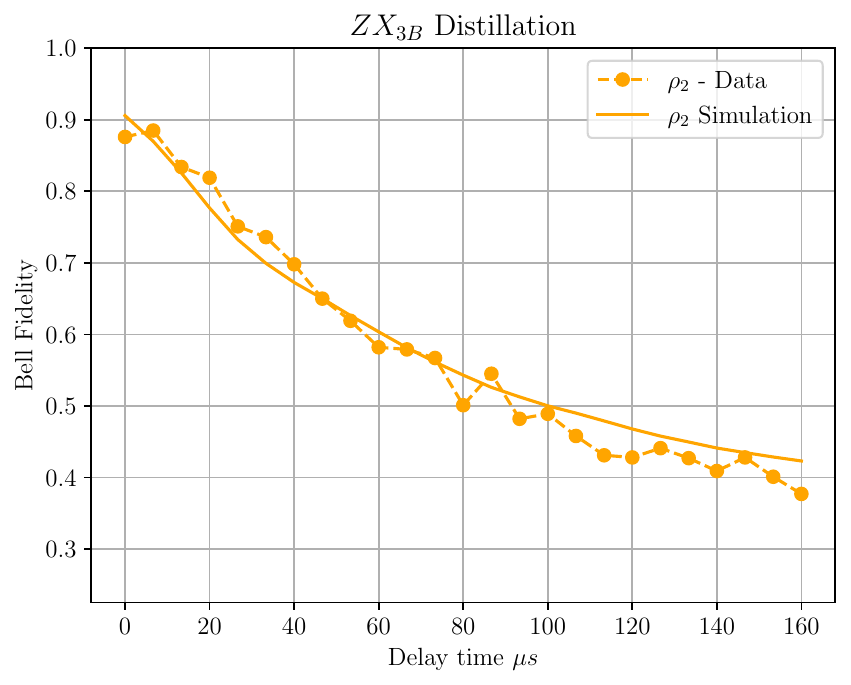}
     \label{fig:3To1Check2Bell2}
     }
    \end{minipage}
    \begin{minipage}{.9\linewidth}
     \subfloat[][Third Bell pair]{
         \includegraphics[clip,width=.4\columnwidth]{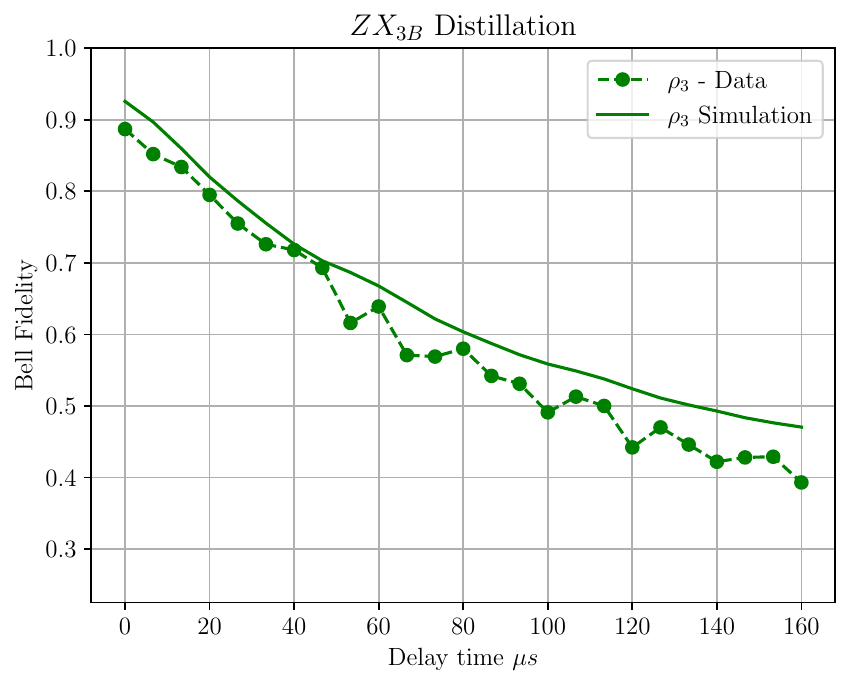}
     \label{fig:3To1Check2Bell3}
     }
     \subfloat[][Distill Bell pair]{
         \includegraphics[clip,width=.4\columnwidth]{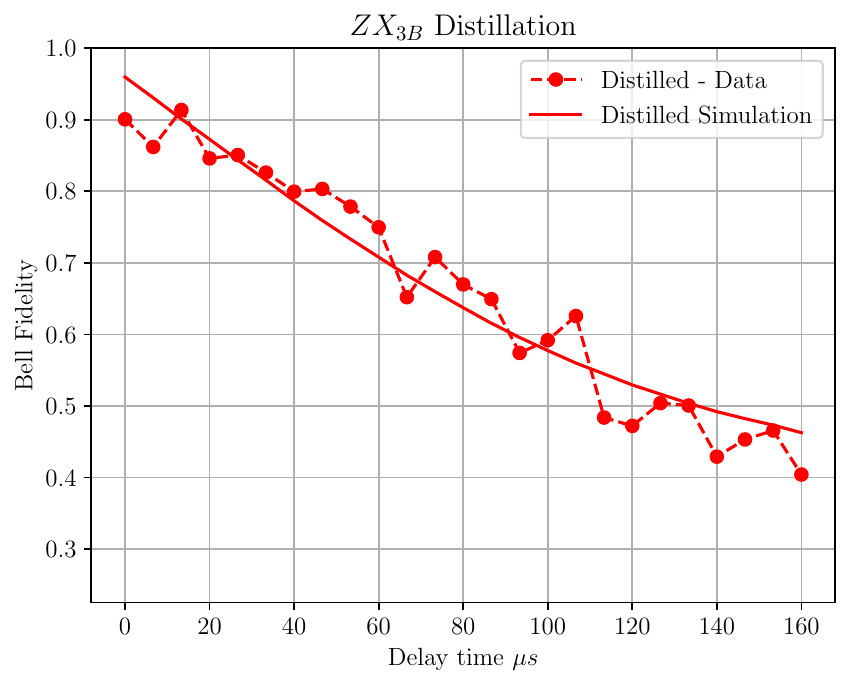}
     \label{fig:3To1Check2Dist}
     }
    \end{minipage}
    \begin{minipage}{.9\linewidth}
     \subfloat[][Acceptance probability]{
         \includegraphics[clip,width=.4\columnwidth]{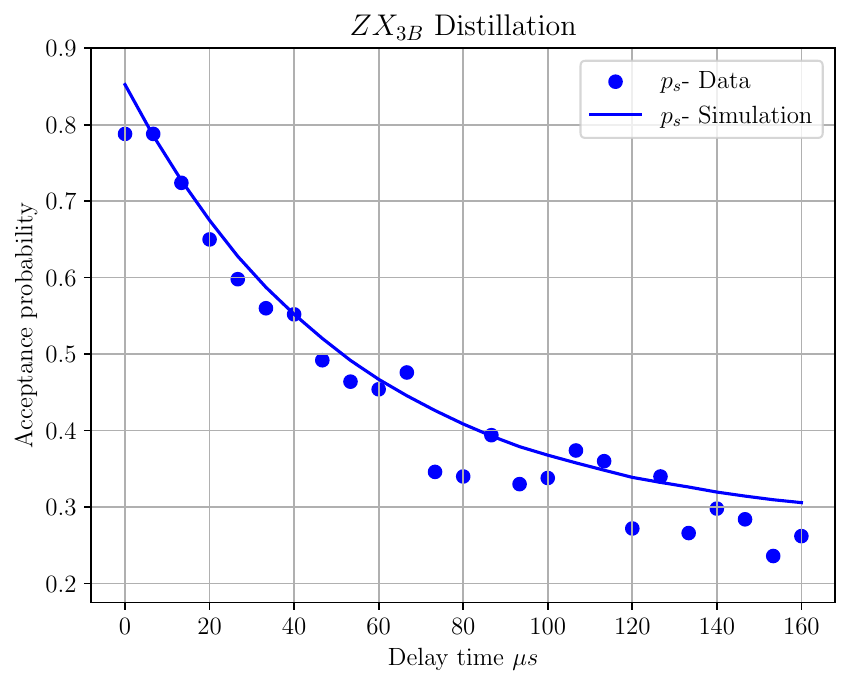}
     \label{fig:3To1Check2Pacc}
     }
    \end{minipage}
     \caption{Bell fidelity as a function of \new{delay} time for pairs before
     and after carrying out $ZX_{3B}$ distillation on qubits $[3, 4, 5, 6, 7,
     8]$ of \texttt{ibm\_kyiv} (points). \new{Solid lines are} simulations
     assuming reported noise parameters as described in
     Table.~\ref{table:3To1DevPar_1} and~\ref{table:3To1DevPar_2} of
     App.~\ref{ap:device}. \new{Figs.~\ref{fig:3To1Check2Bell1},
     ~\ref{fig:3To1Check2Bell2},~\ref{fig:3To1Check2Bell3},
     and~\ref{fig:3To1Check2Dist} plot Bell fidelity against delay time for the
     respective physically non-local state labelled in the figure and
     Fig.~\ref{fig:3To1Check2Pacc} shows the acceptance probability versus
     delay time.} }
     \label{figApp:3To1ExpSim2}
\end{figure*}

\begin{figure*}[htp]
    \begin{minipage}{.9\linewidth} 
    \centering
     \subfloat[][First Bell pair]{
         \includegraphics[clip,width=.4\columnwidth]{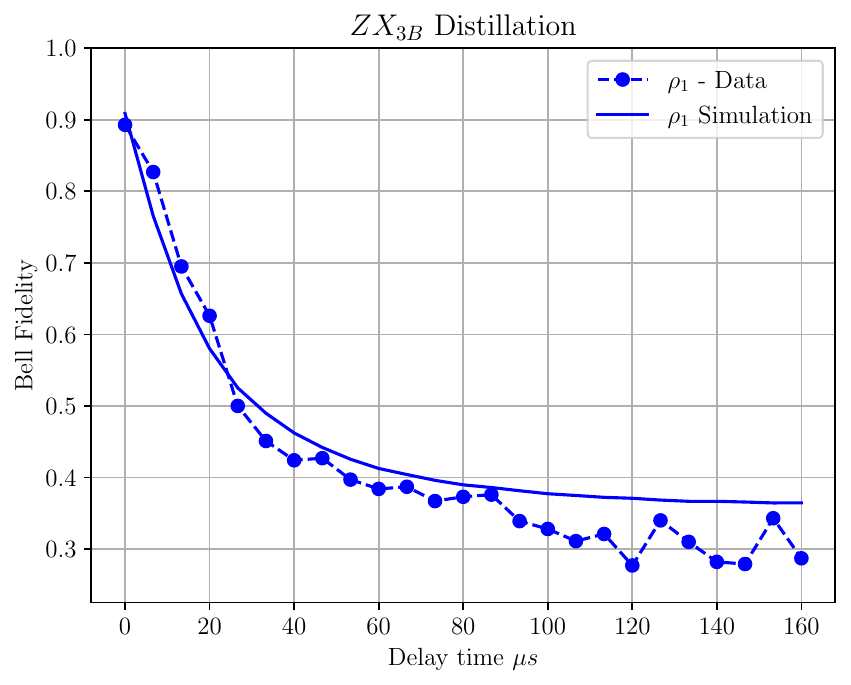}
     \label{fig:3To1CheckBell1}
     }
     \subfloat[][Second Bell pair]{
         \includegraphics[clip,width=.4\columnwidth]{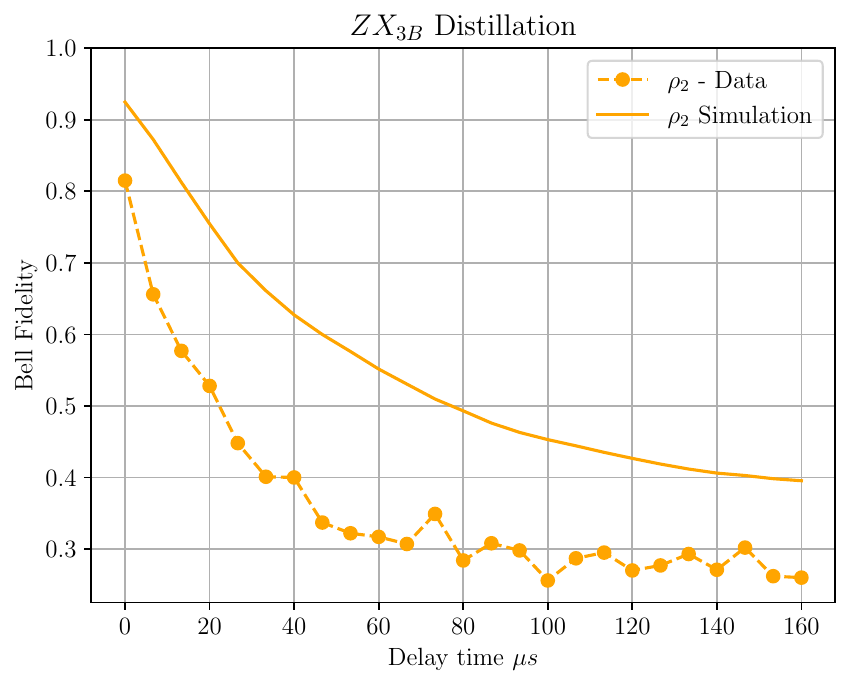}
     \label{fig:3To1CheckBell2}
     }
    \end{minipage}
    \begin{minipage}{.9\linewidth}
     \subfloat[][Third Bell pair]{
         \includegraphics[clip,width=.4\columnwidth]{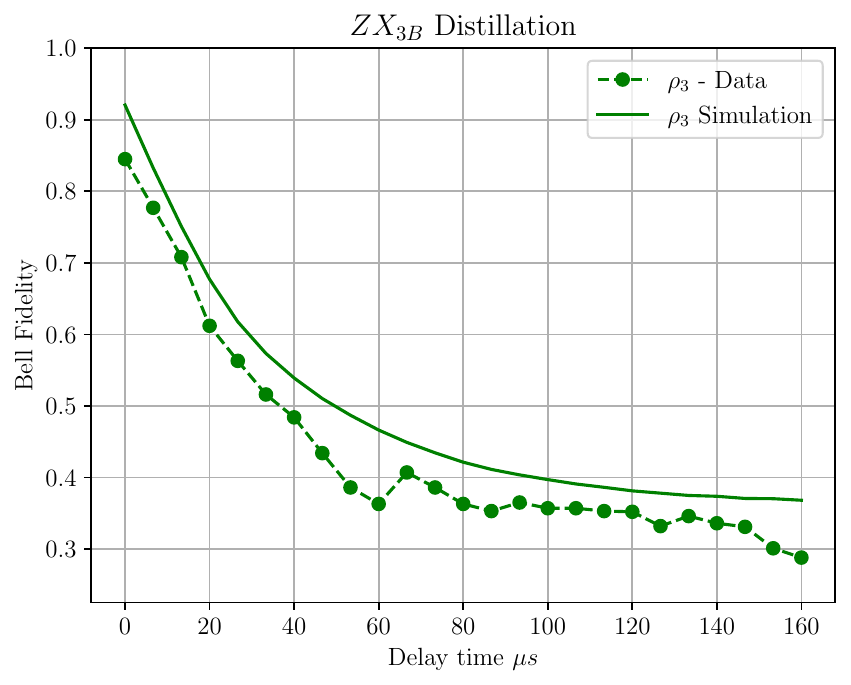}
     \label{fig:3To1CheckBell3}
     }
     \subfloat[][Distill Bell pair]{
         \includegraphics[clip,width=.4\columnwidth]{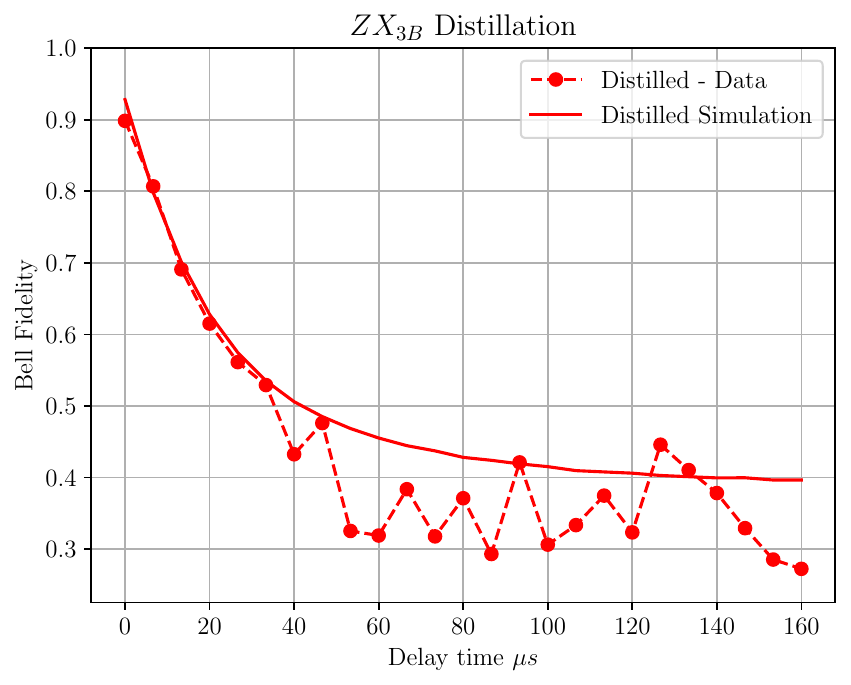}
     \label{fig:3To1CheckBellD}
     }
    \end{minipage}
    \begin{minipage}{.9\linewidth}
     \subfloat[][Acceptance probability]{
         \includegraphics[clip,width=.4\columnwidth]{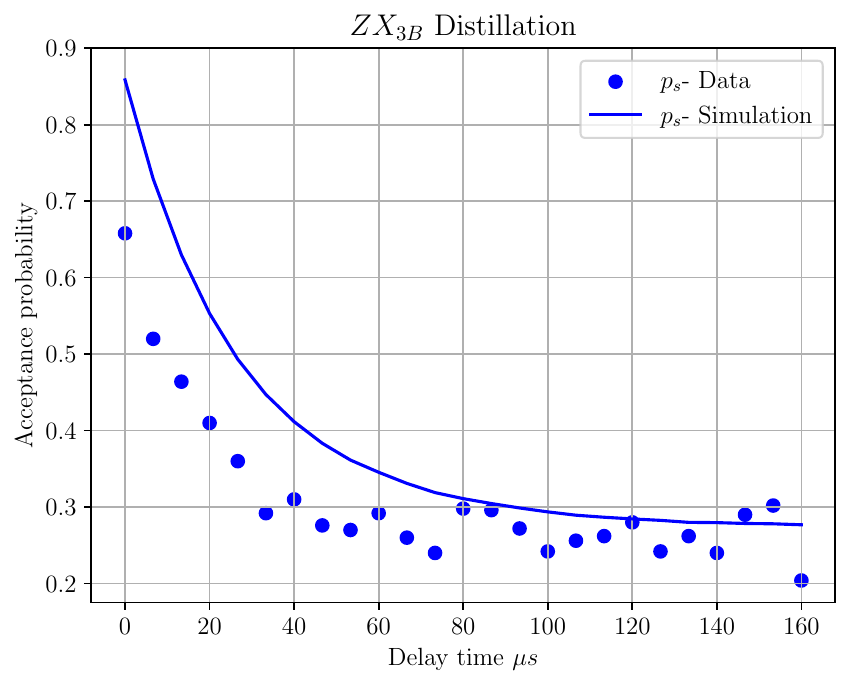}
     \label{fig:3To1CheckPacc}
     }
    \end{minipage}
     \caption{Bell fidelity as a function of \new{delay} time for pairs before
     and after carrying out $ZX_{3B}$ distillation on qubits $[59, 60, 61, 62, 63,
     64]$ of \texttt{ibm\_kyiv} (points). \new{Solid lines are} simulations
     assuming reported noise parameters as described in
     Table.~\ref{table:3To1DevPar_1} and~\ref{table:3To1DevPar_2} of
     App.~\ref{ap:device}. \new{Figs.~\ref{fig:3To1CheckBell1},
     ~\ref{fig:3To1CheckBell2},~\ref{fig:3To1CheckBell3},
     and~\ref{fig:3To1CheckBellD} plot Bell fidelity against delay time for the
     respective physically non-local state labelled in the figure and
     Fig.~\ref{fig:3To1CheckPacc} shows the acceptance probability versus
     delay time.} }
     \label{figApp:3To1ExpSim1}
\end{figure*}

\end{document}